\documentclass[reqno,a4paper]{amsart}

\usepackage{amssymb}
\usepackage{amsmath}

\makeatletter
\@addtoreset{equation}{section}
\makeatother

\renewcommand\thefigure{\thesection.\@arabic\c@figure}
\renewcommand\thetable{\thesection.\@arabic\c@table}
\renewcommand\epsilon{\varepsilon}

\newtheorem{theorem}{Theorem}[section]
\newtheorem{lemma}[theorem]{Lemma}

\newtheorem{corollary}[theorem]{Corollary}

\newtheorem{remark}[theorem]{Remark}

\newcommand{\reff}[1]{(\ref{#1})}

\newcommand{\mc}[1]{{\mathcal #1}}

\newcommand{\mb}[1]{{\mathbf #1}}
\newcommand{\bb}[1]{{\mathbb #1}}

\newcommand{\f}{\varphi}
\newcommand{\e}{\varepsilon}

\newcommand{\rr}{\mathcal{R}}

\def\emptysquare{{\hbox{\vrule height6pt width0.6pt depth0pt%
\vbox{\hrule height0.6pt width4.8pt depth0pt%
\vglue4.8pt%
\hrule height0.6pt width4.8pt depth0pt}%
\vrule height6pt width0.6pt depth0pt}}}

\def\qed{\unskip\nobreak
\hfil\penalty50\hskip1.75em\null\nobreak\hfil\emptysquare
{\parfillskip=0pt \finalhyphendemerits=0 \par}\medskip}

\newenvironment{demo}{\noindent {\bf Proof:}~}{\qed \medskip}

\begin{document}

\begin{titlepage}

\par\vskip 1cm\vskip 2em

\begin{center}

{\LARGE  {\bf
Large deviations for the boundary driven \\ ~\\
symmetric simple exclusion process
}}

\par

\vskip 2.5em \lineskip .5em

{\large
\begin{tabular}[t]{c}
$\mbox{L. Bertini}^{1},\;\, \mbox{ A. De Sole}^{2},\;\,
\mbox{D. Gabrielli}^{3},\;\, \mbox{G. Jona--Lasinio}^{4},\;\, 
\mbox{C. Landim}^{5}$ 
\\
\end{tabular}
\par
}
\medskip
{\small
\begin{tabular}[t]{ll}
{\bf 1} & Dipartimento di Matematica, Universit\`a di Roma La Sapienza\\
&  P.le A.\ Moro 2, 00185 Roma, Italy\\
&  E--mail: {\tt bertini@mat.uniroma1.it}\\
{\bf 2} & Department of Mathematics, MIT \\
& 77 Massachusetts Avenue, Cambridge, MA 02139-4307, USA \\
& E--mail: {\tt desole@math.mit.edu}\\
{\bf 3} & Dipartimento di Matematica, Universit\`a dell'Aquila\\
&  67100 Coppito, L'Aquila, Italy \\
&  E--mail: {\tt gabriell@univaq.it}\\
{\bf 4} & Dipartimento di Fisica and INFN, Universit\`a di Roma La Sapienza\\
&  P.le A.\ Moro 2, 00185 Roma, Italy\\
&  E--mail: {\tt jona@roma1.infn.it}\\ 
{\bf 5}& IMPA, Estrada Dona Castorina 110, J. Botanico, 22460 Rio
de Janeiro, Brazil\\
& CNRS UMR 6085, Universit\'e de Rouen,
76128 Mont--Saint--Aignan Cedex, France \\
& E--mail: {\tt landim@impa.br}\\
\end{tabular}
}
\bigskip
\end{center}

\vskip 1 em
\centerline{\bf Abstract} 
\smallskip
{\small 
\noindent
The large deviation properties of equilibrium (reversible) lattice
gases are mathematically reasonably well understood.  Much less is
known in non--equilibrium, namely for non reversible systems.  In this
paper we consider a simple example of a non--equilibrium situation, the
symmetric simple exclusion process in which we let the system exchange
particles with the boundaries at two different rates. We prove a
dynamical large deviation principle for the empirical density which
describes the probability of fluctuations from the solutions of the
hydrodynamic equation.  The so called quasi potential, which measures
the cost of a fluctuation from the stationary state, is then defined
by a variational problem for the dynamical large deviation rate
function.  By characterizing the optimal path, we prove that the quasi
potential can also be obtained from a static variational problem
introduced by Derrida, Lebowitz, and Speer. 
}

\vfill


%
%





\bigskip\bigskip

\footnoterule
\vskip 1.0em
{\small 
\noindent
Partially supported by Cofinanziamento MURST 2000 and 2001.
\vskip 1.0em
}
\noindent

\end{titlepage}
\vfill\eject

\section{Introduction}

In previous papers \cite{BDGJLlet,BDGJL} we have started the study of
the macroscopic properties of stochastic non equilibrium systems.
Typical examples are stochastic lattice gases which exchange particles
with different reservoirs at the boundary.
In these systems there is a flow of matter through the system and the
dynamics is not reversible.  
The main difference with respect to equilibrium (reversible) states is
the following: in equilibrium the invariant measure, which determines
the thermodynamic properties, is given for free by the Gibbs
distribution specified by the Hamiltonian.  On the contrary, in non
equilibrium states the construction of the appropriate ensemble, that
is the invariant measure, requires the solution of a dynamical
problem.

For equilibrium states, the thermodynamic entropy $S$ is identified 
\cite{C,La,O} with the large deviation rate function for the invariant
measure.  
The rigorous study of large deviations has been extended to
hydrodynamic evolutions of stochastic interacting particle systems
\cite{DV,kov}.  Developing the methods of \cite{kov}, this theory has
been extended to nonlinear hydrodynamic regimes \cite{JLV}.  In a
dynamical setting one may ask new questions, for example what is the
most probable trajectory followed by the system in the spontaneous
emergence of a fluctuation or in its relaxation to equilibrium.  
In the physical literature, the Onsager--Machlup theory \cite{ON2}
gives the following answer under the assumption of time reversibility.
In the situation of a linear macroscopic equation, that is, close to
equilibrium, the most probable emergence and relaxation trajectories
are one the time reversal of the other.

In \cite{BDGJLlet,BDGJL} we have heuristically shown how this theory
has to be modified for non equilibrium systems.  At thermodynamic
level, we do not need all the information carried by the invariant
measure, but only its rate function $S$. This can be obtained, by
solving a variational problem, from the dynamical rate function which
describes the probability of fluctuations from the hydrodynamic
behavior.  The physical content of the variational problem is the
following.  Let $\rho$ be the relevant thermodynamic variable, for
instance the local density, whose stationary value is given by some
function $\bar\rho(u)$. The entropy $S(\rho)$ associated to some
profile $\rho(u)$ is then obtained by minimizing the dynamical rate
function over all possible paths $\pi(t)=\pi(t,u)$ connecting
$\bar\rho$ to $\rho$.  We have shown that the optimal path $\pi^*(t)$
is such that $\pi^*(-t)$ is a solution of the hydrodynamic equation
associated to the time reversed microscopic dynamics, which we call
adjoint hydrodynamics. This relationship is the extension of the
Onsager--Machlup theory to non reversible systems.  Moreover, we have
also shown that $S$ solves an infinite dimensional Hamilton--Jacobi
equation and how the adjoint hydrodynamics can be obtained once $S$ is
known.

\medskip
In the present paper we study rigorously the symmetric one dimensional
exclusion process.  In this model there is at most one particle for
each site of the lattice $\{-N,\dots, N\}$ which can move to a
neighboring site only if this is empty, with rate $1/2$ for each side.
Moreover, a particle at the boundary may leave the system at rate
$1/2$ or enter at rate $\gamma_{-}/2$, respectively $\gamma_+/2$, at
the site $-N$, respectively $+N$. In this situation there is a unique
invariant measure $\mu^N$ which reduces to a Bernoulli measure if
$\gamma_-=\gamma_+$. On the other hand, if $\gamma_-\neq \gamma_+$,
the measure $\mu^N$ exhibits long range correlations \cite{DFIP,S} and
it is not explicitly known.  By using a matrix representation and
combinatorial techniques, Derrida, Lebowitz, and Speer \cite{DLSlet,DLS} have
recently shown that the rate function for $\mu^N$ can be obtained
solving a non linear boundary value problem on the interval $[-1,1]$.
We here analyze the macroscopic dynamical behavior of this system.
The hydrodynamic limit for the empirical density has been proven in
\cite{els1,els2}.  
We prove the associated dynamical large deviation principle which
describes the probability of fluctuations from the solutions of the
hydrodynamic equation. 
We then define the quasi potential via the variational problem
mentioned above. 
By characterizing the optimal path we prove that the quasi potential
can also be obtained from a static variational problem introduced in
\cite{DLSlet,DLS}. 
Using the identification of the quasi potential with the
rate function for the invariant measure proven in \cite{BG},
we finally obtain an independent derivation of the expression 
for the thermodynamic entropy found in \cite{DLSlet,DLS}.

\section{Notation and results}\label{s:nr}
For an integer $N\ge 1$, let $\Lambda_N :=  [-N,N] \cap \bb Z = \{-N
,\dots , N\}$.  The sites of $\Lambda_N$ are denoted by $x$, $y$, and
$z$ while the macroscopic space variable (points in the interval
$[-1,1]$) by $u$, $v$, and $w$. We introduce the microscopic state
space as $\Sigma_N:= \{0,1\}^{\Lambda_N}$ which is endowed with the
discrete topology; elements of $\Sigma_N$, called configurations, are
denoted by $\eta$.  In this way $\eta (x)\in \{0,1\}$ stands
for the number of particles at site $x$ for the configuration $\eta$.

The one dimensional boundary driven simple exclusion process is the
Markov process on the state space $\Sigma_N$ with infinitesimal
generator
$$
L_N \;:=\; L_{-,N} \;+\; L_{0,N} \;+\; L_{+,N} 
$$
defined by 
\begin{eqnarray*}
(L_{0,N} f)(\eta) &:=&  \frac {N^2}2 \sum_{x=-N}^{N-1}  
\big[ f(\sigma^{x,x+1}\eta) - f(\eta) \big]  \; ,\\
(L_{\pm ,N} f)(\eta) &:=&  \frac {N^2}2\,  [\gamma_\pm + (1-\gamma_\pm)
\eta({\pm N})] 
\big[ f(\sigma^{\pm N} \eta) - f(\eta) \big]   
\end{eqnarray*}
for every function $f: \Sigma_N\to\bb R$.  In this formula
$\sigma^{x,y}\eta$ is the configuration obtained from $\eta$ by
exchanging the occupation variables $\eta (x)$ and $\eta (y)$:
$$
(\sigma^{x,y} \eta) (z) \; :=\; \left\{\begin{array}{ll}
                \eta (y) & {\rm if \ } z=x\\
                \eta (x) & {\rm if \ } z=y\\
                \eta (z)   & {\rm if \ } z\neq x,y
                \end{array}
\right.
$$
and $\sigma^{x}\eta$ is the configuration obtained from
$\eta$ by flipping the configuration at $x$:
$$
(\sigma^x \eta)\,(z) \; :=\; \eta (z) [1-\delta_{x,z}] \;+\;
\delta_{x,z} [1-\eta (z)]\; ,
$$
where $\delta_{x,y}$ is the Kronecker delta.  Finally
$\gamma_\pm\in (0,\infty)$ are the activities of the reservoirs at the
boundary of $\Lambda_N$.

Notice that the generators are speeded up by $N^2$; this corresponds
to the diffusive scaling. 
We denote by $\eta_t$ the Markov process on $\Sigma_N$ with generator
$L_N$ and by $\bb P_\eta$ its distribution if the initial
configuration is $\eta$.  Note that $\bb P_\eta$ is a probability
measure on the path space $D(\bb R_+, \Sigma_N)$, which we consider
endowed with the Skorohod topology and the corresponding Borel
$\sigma$--algebra. Expectation with respect to $\bb P_\eta$ is denoted
by $\bb E_\eta$.

\medskip
Our first main result is the dynamical large deviation principle for
the measure $\bb P_\eta$.
We denote by $\langle \cdot,\cdot \rangle$ the inner product in
$L_2\big([-1,1],du\big)$ and let 
\begin{equation}\label{dcm}
\mc M := \left\{ \rho \in L_\infty \big([-1,1],du\big) \,:\:
0\le \rho (u) \le 1 \: \textrm{ a.e.} \right\}
\end{equation}
which we equip with the topology induced by weak convergence, namely
$\rho_n \to \rho$ in $\mc M$ if and only if $\langle \rho_n, G \rangle
\to \langle \rho, G \rangle$ for each continuous function $G:
[-1,1]\to\bb R$; we consider $\mc M$ also endowed with the
corresponding Borel $\sigma$--algebra.  Let us define the map $\pi^N
\colon\Sigma_N \to \mc M$ as
\begin{equation}
\label{eq:2}
\pi^N (\eta) \;:=\; 
\sum_{x=-N}^{N} \eta (x) 
\mb 1{\left\{ \Big[ \frac{x}{N}- \frac{1}{2N}, 
\frac{x}{N}+ \frac{1}{2N}\Big)\right\}} \; , 
\end{equation}
where $\mb 1\{ A\}$ stands for the indicator function of the set $A$;
namely $\pi^N =\pi^N(\eta)$ is the empirical density obtained from the
configuration $\eta$.  Notice that $\pi^N(\eta)\in\mc M$, i.e.\ $0\le
\pi^N(\eta)\le 1$, because $\eta (x)\in\{0,1\}$.

Let $\eta^N$ be a sequence of configurations for which the empirical
density $\pi^N(\eta^N)$ converges in $\mc M$, as $N\uparrow\infty$, to
some function $\rho$, namely for each $G\in C\big([-1,1]\big)$  
\begin{equation}
\label{f01}
\lim_{N\to\infty} \langle \pi^N(\eta^N), G \rangle = 
\lim_{N\to\infty} \sum_{x=-N}^{N} \eta^N(x) 
\int_{(-1) \vee \big(\frac xN - \frac {1}{2N}\big) }
     ^{1\wedge \big(\frac xN + \frac {1}{2N}\big)}
\!\!\!\!\!\!\!du \: G(u)
= \int_{-1}^1 \!du \: \rho (u) G(u)
\end{equation}
where we used the notation $a\wedge b
:= \min\{a,b\}$ and $a\vee b := \max\{a,b\}$.  If \reff{f01} holds we
say that the sequence $\{\eta^N: N\ge 1\}$ is associated to the
profile $\rho \in\mc M$.

For $T>0$ and positive integers $m$, $n$ we denote by $C^{m,n}_0
([0,T]\times [-1,1])$ the space of functions $G\colon [0,T]\times
[-1,1] \to\bb R$ with $m$ continuous derivatives in time, $n$
continuous derivatives in space and which vanish at the boundary:
$G(\cdot , \pm 1)=0$.  Let also $D\big([0,T], \mc M\big)$ be the
Skorohod space of paths from $[0,T]$ to $\mc M$ equipped with its
Borel $\sigma$--algebra. Elements of $D\big([0,T], \mc M\big)$ will be
denoted by $\pi(t)=\pi(t,u)$.

Let $\rho_\pm := \gamma_\pm /[1+\gamma_\pm]\in(0,1)$ be the density at
the boundary of $[-1,1]$ and fix a function $\rho\in\mc
M$ which corresponds to the initial profile.  For $H\in
C^{1,2}_0([0,T]\times [-1,1])$, let $J_{T, H, \rho} = J_{H}\colon
D([0,T], \mc M)\longrightarrow \bb R$ be the functional given by
\begin{eqnarray}\label{j}
J_H(\pi) &:=& \big\langle \pi(T), H(T) \big\rangle 
- \langle \rho, H(0)\rangle
- \int_0^{T} \!dt\, \Big\langle \pi(t), \partial_t H(t) 
+ \frac 12 \Delta H(t) \Big\rangle
\nonumber\\
&& + \frac{\rho_+}{2}   \int_0^{T} \! dt\, \nabla H(t,1) 
\; -\;  \frac{\rho_-}{2} \int_0^{T} \!dt\, \nabla H(t,-1) \nonumber\\
&& -\frac{1}{2} \int_0^{T} \!dt\, 
\big\langle \chi( \pi(t) ), \big( \nabla H(t) \big)^2 \big\rangle \; ,
\end{eqnarray}
where $\nabla$ denotes the derivative with respect to the macroscopic
space variable $u$, $\Delta$ is the Laplacian on $(-1,1)$, and we have
set
$\chi(a) := a(1-a)$.  
Let finally $I_T (\,
\cdot \, | \rho) \colon D([0,T],\mc M)\longrightarrow [0,+\infty]$ be
the functional defined by
\begin{equation}
\label{eq:3}
I_T (\pi | \rho) \; :=\; \sup_{H\in C^{1,2}_0([0,T]\times [-1,1])}
J_H(\pi)\; .
\end{equation}
Notice that, if $\pi(t)$ solves the heat equation with boundary condition
$\pi(t,\pm1)=\rho_\pm$ and initial datum $\pi(0)=\rho$, then $I_T(\pi|\rho)=0$.

\begin{theorem}
\label{s02}
Fix $T>0$ and a profile $\rho\in\mc M$ bounded away from $0$ and $1$,
namely such that there exists $\delta>0$ with $\delta \le \rho \le
1-\delta$ a.e.  
Consider a sequence $\eta^N$ of configurations associated to
$\rho$. Then the measure $\bb P_{\eta^N} \circ (\pi^N)^{-1}$ on
$D\big([0,T],\mc M\big)$ satisfies a large deviation principle with
speed $N$ and convex lower semi continuous rate function
$I_T(\cdot|\rho)$. Namely, for each closed set $\mc C \subset D([0,T],
\mc M)$ and each open set $\mc O \subset D([0,T], \mc M)$,
\begin{equation*}
\begin{array}{l}
{\displaystyle 
\limsup_{N\to\infty} \frac 1N \log \bb P_{\eta^N} [ \pi^N \in \mc C]
\;\le\; - \inf_{\pi \in \mc C} I_T (\pi | \rho)  
}\\
\qquad {\displaystyle 
\liminf_{N\to\infty} \frac 1N \log \bb P_{\eta^N} [ \pi^N\in
\mc O] \;\ge\; - \inf_{\pi \in \mc O} I_T (\pi | \rho) \; .
}
\end{array}
\end{equation*}
\end{theorem}

It is possible to obtain a more explicit representation of
the functional $I_T (\cdot | \rho)$, see Lemma \ref{s07} below.  If
the particle system is considered with periodic boundary conditions,
i.e.\ $\Lambda_N$ is replaced by the discrete torus of length $N$,
this Theorem has been proven in \cite{kov}.  As we shall see later,
the main difference with respect to the case with periodic boundary
condition is the lack of translation invariance and the fact that the
path $\pi(t,\cdot)$ is fixed at the boundary.

\medskip
We now define precisely the variational problem mentioned in the
introduction.  Let $\bar\rho\in \mc M$ be the linear profile
$\bar\rho(u) := \left[ \rho_- (1-u) + \rho_+ (1+u)\right]/2$, $u\in
[-1,1]$, which is the density profile associate to the invariant
measure $\mu^N$, see Section \ref{s:3} below.  We then define $V:
\mc M \to \bb [0,+\infty]$ as the quasi potential for the rate function
$I_T(\,\cdot\,|\bar\rho)$:
\begin{equation}
\label{qp}
V(\rho) \;:=\; \inf_{T>0}\; 
\inf_{\pi(\cdot) \::\: \pi(T)=\rho} \; I_T(\pi | \bar\rho) \; .
\end{equation}
which measures the minimal cost to produce the profile $\rho$
starting from $\bar\rho$.

Let us first describe how the variational problem \reff{qp} is solved
when $\gamma_-=\gamma_+ =\gamma$.  In this case
$\bar\rho=\gamma/(1+\gamma)$ is constant and the process is reversible
with respect to the Bernoulli measure with density $\bar\rho$.
We have that $I_T(\pi |\rho_0)= 0$ if $\pi(t)$ solves the hydrodynamic
equation which for this system is given by the heat equation:
\begin{equation}\label{hrho}
\left\{
\begin{array}{l}
{\displaystyle \partial_t \rho(t) = (1/2) \Delta \rho(t)}
\; , \\
{\displaystyle \rho(t, \pm 1) = \rho_\pm }\; , \\
{\displaystyle \rho(0, \cdot) =  \rho_0 (\cdot) } \; .
  \end{array}
\right.
\end{equation}
Note that $\rho(t)\rightarrow \bar\rho$ as $t\to \infty$. 

It can be easily shown that the minimizer for \reff{qp}, defined on
the time interval $(-\infty, 0]$ instead of $[0,+\infty)$ as in
\reff{qp}, is given by $\pi^*(t)=\rho (-t)$, where $\rho(t)$ is the
solution of \reff{hrho} with initial condition $\rho_0=\rho$.
This symmetry of the relaxation and fluctuation trajectories is
the Onsager--Machlup principle mentioned before.

Moreover the quasi potential $V(\rho)$ coincides with the entropy of
the Bernoulli measure with density $\bar\rho$, that is, 
understanding $0\log 0 =0$, 
\begin{equation}\label{S_0}
V(\rho) = S_0 (\rho) := \int_{-1}^1 \!du \: 
\Big[ \rho (u) \log \frac {\rho
(u)}{\bar\rho} + [1 - \rho(u)] \log \frac {1- \rho(u)}
{1-\bar\rho}\Big] 
\end{equation}

\medskip
In the context of Freidlin--Wentzell theory \cite{FW} 
for diffusions in $\bb
R^n$, the situation just described is analogous to the so called {\em
gradient\/} case in which the quasi potential coincides with the
potential. This structure reflects the reversibility of the underlying
process. 
In general for {\em non gradient\/} systems, the solution of the dynamical
variational problem, or of the associated Hamilton--Jacobi equation,
cannot be explicitly calculated. 
The case $\gamma_+\neq\gamma_-$ is analogous to a non gradient system,
but for this particular model we shall prove that the 
quasi potential $V(\rho)$, as defined in \reff{qp}, coincides with the
functional $S(\rho)$ defined by a time independent variational problem
introduced in \cite{DLSlet,DLS} which is stated below.
This is the second main result of this paper.

Denote by $C^1\big([-1,1]\big)$ the space of once continuously
differentiable functions $f:[-1,1]\to \bb R$ endowed with the norm
$\|f\|_{C^1} := \sup_{u\in[-1,1]} \big\{ |f(u)| + |f'(u)| \big\}$.
Let
\begin{equation}
\label{dF}
\mc {F} :=\Big\{f \in  C^1\big([-1,1]\big) \,:\: f(\pm 1)
=\rho_\pm,\;\;  [\rho_+-\rho_-] f'(u) > 0 \, , u\in [-1,1] \Big\} \; ,
\end{equation}
where $f'$ denotes the derivative of $f$. Note that if $f\in\mc F$
then $0<\rho_-\wedge\rho_+ \le f(u) \le \rho_-\vee\rho_+ <1$ for all
$-1\le u\le 1$.

For $\rho\in \mc M$ and $f\in \mc F$ we set
\begin{equation}
\label{dG}
\mc G (\rho, f) :=  
\int_{-1}^1 \!du\, \Big[ \rho (u) \log \frac {\rho
(u)}{f(u)} + \big[1 - \rho (u)\big] \log \frac {1- \rho
(u)}{1- f(u)} + \log \frac {f'(u)}{[\rho_+ - \rho_-]/2}
\Big]
\end{equation} 
and
\begin{equation}
\label{SsS}
S(\rho) := \sup_{f\in\mc F} \; \mc G (\rho, f)\; .
\end{equation}
Theorem \ref{t:s=s} below, which formalizes the arguments in
\cite{DLS}, states that the supremum in \reff{SsS} is uniquely
attained for a function $f$ which solves a non linear boundary value
problem. We shall denote it by $F=F (\rho)$ to emphasize its
dependence on $\rho$; therefore $S(\rho) = \mc G \big( \rho, F(\rho)
\big)$.

\begin{theorem}
\label{qp=s}
Let $V$ and $S$ as defined in \reff{qp} and \reff{SsS}. Then
for each $\rho \in \mc M$  we have $V(\rho)=S(\rho)$.
\end{theorem}

In the proof of the above theorem we shall construct a particular path
$\pi^*(t)$ in which the infimum in (\ref{qp}) is almost attained. As
recalled in the introduction, by the heuristic arguments in \cite{BDGJL}, 
$\pi^*(-t)$ is the solution 
of the hydrodynamic equation corresponding to the process with generator
$L_N^*$, the adjoint of $L_N$ in $L_2(\Sigma_N,d\mu^N)$ and initial
condition $\rho$.
In analogy to the Freidlin--Wentzell theory \cite{FW}, we expect that
the exit path from a neighborhood $\bar\rho$ to a neighborhood of
$\rho$ should, with probability converging to one as
$N\uparrow\infty$, take place in a small tube around the path $\pi^*(t)$.

The optimal path can be described in a rather simple
fashion. Recalling that we denoted by $F=F(\rho)$ the maximizer
for \reff{SsS}, consider the heat equation in $[-1,1]$ with 
boundary conditions $\rho_\pm$ and initial datum $F$:
\begin{equation}\label{hF}
\left\{
\begin{array}{l}
{\displaystyle \partial_t \Phi(t) = (1/2) \Delta \Phi(t)}
\; , \\
{\displaystyle \Phi(t, \pm 1) = \rho_\pm }\; , \\
{\displaystyle \Phi(0, \cdot) =  F( \rho )} \; .
  \end{array}
\right.
\end{equation}
We next define $\rho^*(t)=\rho^*(t,u)$ by
\begin{equation}
\label{eq:4}
\rho^*(t) \;:=\; \Phi(t) \; +\;  \Phi(t)[1-\Phi(t)] 
\frac {\Delta \Phi(t)}{ \big(\nabla \Phi(t)\big)^2} \; \cdot  
\end{equation}
In view of (\ref{Deq}) below, $\rho^*(0)=\rho$ and, by Lemma
\ref{convergence!}, $\lim_{t\to\infty}\rho^*(t)=\bar\rho$.  The
optimal path $\pi^*(t)$, defined on the time interval $(-\infty, 0]$
instead of $[0,+\infty)$ as in \reff{qp}, is then given by
$\pi^*(t)=\rho^*(-t)$.

\medskip From the dynamical large deviation principle we can obtain,
by means of the quasi potential, the large deviation principle for
the empirical density when the particles are distributed according to
the invariant measure of the process $\eta_t$.
Note that the finite state Markov process $\eta_t$ with generator
$L_N$ is irreducible, therefore it has a unique invariant measure 
$\mu^N$. 

Let us introduce $\mc P_N := \mu^N \circ \big( \pi^N \big)^{-1}$ which
is a probability on $\mc M$ and describes the behavior of the
empirical density under the invariant measure.  In
\cite{DFIP,els1,els2,S} it is shown, see also Section \ref{s:3}
below, that $\mc P_N$ satisfies the law of large numbers $\mc P_N
\Rightarrow \delta_{\bar\rho}$ in which $\Rightarrow$ stands for weak
convergence of measures on $\mc M$ and $\bar\rho$ is the linear
profile already introduced.  

Since $\bar\rho$ is globally attractive for \reff{hrho}, 
the quasi potential with
respect to $\bar\rho$ defined in \reff{qp} gives the rate
function for the family $\mc P_N$. In \cite{BDGJLlet,BDGJL} we have
heuristically derived this identification via a time reversal
argument. For the present model a rigorous proof, in the same
spirit of the Freidlin--Wentzell theory, is 
given in \cite{BG}; that is we have the following theorem.

\begin{theorem}
\label{t:BG}
Let $V$ as defined in \reff{qp}. Then  the measure $\mc P_N$ satisfies
a large deviation principle with speed $N$ and rate function $V$.
\end{theorem}

The identification of the rate function for $\mc P_N$ with the
functional $S$ now follows from Theorems \ref{s02}, \ref{qp=s} and
\ref{t:BG}.
 
\begin{corollary}\label{t:ldp}
Let $S$ as defined in \reff{SsS}. The measure $\mc P_N$ satisfies a
large deviation principle on $\mc M$ with speed $N$ and convex lower
semi continuous rate function $S$. Namely for each closed set $\mc
C\subset\mc M$ and each open set $\mc O\subset \mc M$,
\begin{equation*}
\begin{array}{l}
{\displaystyle 
\limsup_{N\to\infty} \frac 1N \log \, \mc \mu^N [ \pi^N \in \mc C]
\le - \inf_{\rho \in \mc C} S (\rho)  
}\\
\quad {\displaystyle 
\liminf_{N\to\infty} \frac 1N \log \, \mu^N [ \pi^N \in \mc O]
\ge- \inf_{\rho \in \mc O} S(\rho) 
}\end{array}
\end{equation*}
\end{corollary}

As already mentioned, the rate function $S$ has been first obtained in
\cite{DLSlet,DLS} by using a matrix representation of the invariant
measure $\mu^N$ and combinatorial techniques. By means of Theorems
\ref{s02}, \ref{qp=s}, and \ref{t:BG} we prove here, independently of
\cite{DLSlet,DLS}, the large deviation principle by following the
dynamical/variational route explained in \cite{BDGJL} which is
analogous to the Freidlin--Wentzell theory \cite{FW} for diffusions on
${\bb R}^n$.

\smallskip
We remark that it should be possible, modulo technical problems, to
extend Theorems \ref{s02} and \ref{t:BG} to other boundary driven
diffusive lattice gases, see \cite{BDGJL} for a heuristic discussion.
The characterization of the rate function for the invariant measure as
the quasi potential allows to obtain some information on it directly
from the variational problem \reff{qp}. In particular, in Appendix
\ref{s:app}, we discuss the symmetric simple exclusion in any
dimension and get a lower bound on $V$ in terms of the entropy $S_0$
of the equilibrium system. In the one dimensional case, this bound has
been proven in \cite{DLSlet,DLS} by using instead the variational problem
\reff{SsS}.

\medskip
\noindent{\it Outline.}\
In Section \ref{s:3} we recall the hydrodynamic behavior of the
boundary driven exclusion process and prove the associated large
deviation principle described by Theorem \ref{s02}. In Section
\ref{s:4} and \ref{s:5}, which are more technical, we state and prove
some properties of the functional $S$ which is then shown to coincide
with the quasi potential $V$. Finally, in Appendix \ref{s:app}, we
consider the symmetric simple exclusion in any dimension and prove a
lower bound on $V$.

\section{Dynamical behavior}
\label{s:3}
In this section we study the dynamical properties of the empirical
density for the boundary driven simple exclusion process in a fixed
(macroscopic) time interval $[0,T]$.  In particular we review the
hydrodynamic limit (law of large numbers) and prove the
corresponding large deviation principle.  This problem was considered
before by Kipnis, Olla and Varadhan in \cite{kov} for the exclusion
process with periodic boundary condition. For this reason, we present
only the modifications needed in the argument and refer to
\cite{kov,kl,bkl} for the missing arguments. 

As already stated, the invariant measure $\mu^N$ is not known
explicitly but some of its properties have been derived. For example,
the one site marginals or the correlations can be computed explicitly.
To compute the one site marginals, which will be used later, let
$\rho^N(x) = E_{\mu^N}[\eta(x)]$ for $-N \le x\le N$. Since $\mu^N$ is
invariant, $E_{\mu^N}[L_N \eta(x)] = 0$. Computing $L_N \eta(x)$, we
obtain a closed difference equation for $\rho^N(x)$:
$$
\left\{
  \begin{array}{l}
\vphantom{\Big\{}
(\Delta_N \rho^N)(x) = 0 \quad \text{for $-N+1 \le x\le N-1$}\; , \\
\vphantom{\Big\{}
\rho^N (N-1) - \rho^N (N) + \gamma_+ [1-\rho^N (N)] - \rho^N (N)  =0 \; , \\
\vphantom{\Big\{}
\rho^N (-N+1) - \rho^N (-N) + \gamma_- [1-\rho^N (-N)] - \rho^N (-N)  =0 \; .
  \end{array}
\right.
$$
In this formula, $\Delta_N$ stands for the discrete Laplacian so
that $(\Delta_N f)(x) = f(x+1) + f(x-1) - 2 f(x)$. The unique solution
of this discrete elliptic equation gives the one--site marginals of $\mu^N$.

Denote by $\nu^N = \nu^N_{\gamma_-, \gamma_+}$ the product measure on
$\Sigma_N$ with marginals given by
$$
\nu^N \{\eta : \eta(x) =1\} \; =\; \rho^N(x)
$$
and observe that the generators $L_{-,N}$, $L_{+,N}$ are reversible
with respect to $\nu^N$.

Denote by $\{\tau_x : x\in \bb Z\}$ the group of translations in
$\{0,1\}^{\bb Z}$ so that $(\tau_x \zeta) (z) = \zeta (x+z)$ for all
$x$, $z$ in $\bb Z$ and configuration $\zeta$ in $\{0,1\}^{\bb Z}$.
Translations are extended to functions and measures in a natural way.
Eyink, Lebowitz and Spohn \cite{els1} and De Masi, Ferrari, Ianiro and
Presutti \cite{DFIP} proved that
$$
\lim_{N\to \infty} E_{\mu^N} [\tau_{[uN]} f] \; =\; 
E_{\nu_{\bar \rho (u)}} [f]
$$
for every local function $f$ and $u$ in $(-1,1)$. Here $\bar\rho$
is the unique solution of 
$$
\left\{
  \begin{array}{l}
(1/2) \Delta \rho = 0 \; , \\
\rho(\pm1) = \rho_\pm \; ,
  \end{array}
\right.
$$
namely $\bar\rho$ is the linear interpolation between $\rho_-$ and
$\rho_+$ and 
$\{\nu_\alpha : 0\le \alpha\le 1\}$ stands for the Bernoulli product
measure in $\{0,1\}^{\bb Z}$ with density $\alpha$ and $\rho_\pm =
\gamma_\pm /[1+\gamma_\pm]$ is the density at the boundary of
$[-1,1]$. 

\subsection{Hydrodynamic limit}
\label{s:3.1}
Recall that, for each configuration $\eta\in\Sigma_N$, we denote by
$\pi^N = \pi^N(\eta)\in \mc M$ the empirical density obtained from
$\eta$, see equation \reff{eq:2}. We say that a sequence of
configurations $\{\eta^N: N\ge 1\}$ is associated to the profile
$\gamma$ if \reff{f01} holds for all continuous functions $G:[-1,1]\to
\bb R$.  The following result is due to Eyink, Lebowitz and Spohn
\cite{els2}.

\begin{theorem}
\label{s01}
Consider a sequence $\eta^N$ associated to some profile $\rho_0\in\mc
 M$. Then, for all $t>0$, $\pi^N(t) = \pi^N(\eta_{t})$
converges (in the sense \reff{f01}) in probability to $\rho(t,u)$,
the unique weak solution of
\begin{equation}
\label{f02}
\left\{
  \begin{array}{l}
\partial_t \rho = (1/2) \Delta \rho \; , \\
\rho(t, \pm1) = \rho_\pm \; , \\
\rho(0, \cdot) = \rho_0(\cdot)\; .
  \end{array}
\right.
\end{equation}
\end{theorem}

By a weak solution of the Dirichlet problem \reff{f02} in the time
interval $[0,T]$, we understand a {\it bounded} real function $\rho$
which satisfies the following two conditions.
\begin{description}
\item [(a)] There exists a function $A(t,u)$ in $L^2([-1,1]\times
[0,T])$ such that 
\begin{eqnarray*}
&& \int_0^t ds \int_{-1}^1 du \, \rho(s,u) ( \nabla H)(u) \\
&& \qquad\qquad\quad
=\; \{\rho_+ H(1) - \rho_- H(-1) \} t
\; -\;  \int_0^t ds \int_{-1}^1 du \, A(s,u) H(u) 
\end{eqnarray*}
for every smooth function $H\colon [-1,1]\to \bb R$ and every $0\le
t\le T$. $A(t,u)$ will be denoted by $(\nabla \rho)(t,u)$.

\item[(b)] For every function $H\colon [-1,1]\to \bb R$ of class
$C^1\big([-1,1]\big)$ vanishing at the boundary and every $0\le t\le T$,
\begin{eqnarray*}
&& \int_{-1}^1 du \, \rho(t,u) H(u) \; -\; 
\int_{-1}^1 du \, \rho_0(u) H(u) \\
&& \qquad\qquad\qquad\qquad
=\; - (1/2) \int_0^t ds \int_{-1}^1 du \, ( \nabla \rho)(s,u) 
( \nabla H)(u) \; .
\end{eqnarray*}
\end{description}
 
The classical $H_{-1}$ estimates gives uniqueness of weak solutions of
equation \reff{f02}. Note that here the weak solution coincides with
the semi--group solution $\rho (t) = \bar \rho + e^{t
\Delta^0/2} (\rho_0 - \bar \rho)$, where $\bar\rho$ is the stationary
profile and $\Delta^0$ is the Laplacian with zero boundary condition.

\subsection{Super-exponential estimate}
We now turn to the problem of large deviations from the hydrodynamic
limit.  It is well known that one of the main steps in the derivation
of a large deviation principle for the empirical density is a
super--exponential estimate which allows the replacement of local
functions by functionals of the empirical density in the large
deviations regime. Essentially, the problem consists in bounding
expressions such as $\langle V, f^2\rangle_{\mu^N}$ in terms of the
Dirichlet form $\langle-L_N f, f\rangle_{\mu^N}$.
Here $V$ is a local function and $\langle \cdot, \cdot
\rangle_{\mu^N}$ indicates the inner product with respect to the
invariant state $\mu^N$.

In the context of boundary driven simple exclusion processes, the fact
that the invariant state is not known explicitly introduces a
technical difficulty.  Following \cite{lov1} we fix $\nu^N$, the
product measure defined in the beginning of this section, as reference
measure and estimate everything with respect to $\nu^N$. However,
since $\nu^N$ is not an invariant state, there are no reasons for
$\langle -L_N f, f\rangle_{\nu^N}$ to be positive. The first statement
shows that this expression is almost positive.

For a function $f\colon \Sigma_N\to \bb R$, let
$$
D_N(f) \;=\; \sum_{x=-N}^{N-1} \int [ f(\sigma^{x,x+1} \eta) -
f(\eta)]^2 \, d\nu^N(\eta) \; .
$$

\begin{lemma}
\label{s03}
There exists a finite constant $C_0$ depending only on $\gamma_\pm$
such that
$$
\langle L_{0,N} f, f\rangle_{\nu^N} \;\le \; - \frac{N^2}4 D_N(f) \;+\;
C_0 N \langle f, f\rangle_{\nu^N} 
$$
for all functions $f\colon \Sigma_N\to \bb R$
\end{lemma}

The proof of this lemma is elementary and left to the reader. Notice
on the other hand that both $\langle -L_{+,N} f, f\rangle_{\nu^N}$,
$\langle-L_{-,N} f, f\rangle_{\nu^N}$ are positive because $\nu^N$ is
a reversible state by our choice of the profile $\rho^N$.

This lemma together with the computation presented in \cite[p.
78]{bkl} for non--reversible processes, permits to prove the
super--exponential estimate. The statement of this result requires
some notation.  For a cylinder function $\Psi$, denote the expectation
of $\Psi$ with respect to the Bernoulli product measure $\nu_\alpha$
by $\tilde\Psi(\alpha)$:
$$
\tilde\Psi(\alpha) \;:=\; E_{\nu_\alpha}[\Psi]\; .
$$
For a positive integer $\ell$ and $-N \le x \le N$, denote the
empirical mean density on a box of size $2\ell +1$ centered at $x$ by
$\eta^\ell (x)$, namely
$$
\eta^\ell(x) \;=\; \frac 1{|\Lambda_\ell (x)|}
\sum_{y\in \Lambda_\ell (x)} \eta(y)\; ,
$$
where $\Lambda_\ell (x) = \Lambda_{N,\ell} (x) = \{y\in \Lambda_N
:\, |y-x|\le \ell\}$.  Let $H\in C([0,T]\times [-1,1])$ and $\Psi$ a
cylinder function.  For $\epsilon>0$, define also
$$
V_{N,\epsilon}^{H,\Psi}(t,\eta) \;=\; \frac 1N \sum_{x} H(t,x/N)
\Big\{ \tau_x\Psi(\eta)-
\tilde\Psi\big(\eta^{N\epsilon}(x)\big)\Big\}\; ,
$$
where the summation is carried over all $x$ such that the
support of $\tau_x \Psi$ belongs to $\Lambda_N$. For a continuous
function $G: [0,T]\to \bb R$, let 
$$
W_G^\pm \;=\; \int_0^T ds\, G(s) [ \eta_s(\pm N) - \rho_{\pm} ]\; .
$$

\begin{theorem}
\label{s04}
Fix $H$ in $C([0,T]\times [-1,1])$, $G\in C([0,T])$, a cylinder 
function $\Psi$, and a sequence $\{\eta^N \in
\Sigma_N : N\ge 1\}$ of configurations.  For any $\delta>0$ we have
$$
\limsup_{\epsilon\to 0}\,\limsup_{N\to\infty} \frac 1N  \log
\bb P_{\eta^N} \Big [\, \Big |
\int_0^{T}V_{N,\epsilon}^{H,\Psi}(t,\eta_t)\, dt \Big | >\delta
\Big] \; =\; -\infty \; , 
$$
$$
\limsup_{N\to\infty} \frac 1N  \log
\bb P_{\eta^N} \Big [\, \big | W_G^\pm \big | > \delta \Big] \; =\;
-\infty  \; .
$$
\end{theorem}

\subsection{Upper bound}
The proof of the upper bound of the large deviation principle is
essentially the same as in \cite{kov}. There is just a slight
difference in the definition of the functionals $J_H$ due to the
boundary conditions.

For $H$ in $C^{1,2}_0([0,T]\times [-1,1])$ consider the exponential
martingale $M^H_t$ defined by
$$
\begin{array}{rcl}
{\displaystyle 
M^H_t 
}
&=&  
{\displaystyle 
\exp \bigg\{ N \Big[
\langle\pi^N(t), H(t)\rangle - \langle\pi^N(0), H(0)\rangle 
}
\\
&&
{\displaystyle 
\phantom{\exp N \Big \{} 
\; -\;
\frac 1N \int_0^t e^{- N \langle\pi^N(s), H(s)\rangle} 
(\partial_s + N^2 L_N) 
e^{N\langle\pi^N(s), H(s)\rangle} \,ds \Big] \bigg\}\; .
}
\end{array}
$$
An elementary computation shows that
$$
M^H_T \;=\;  \exp N\Big\{  J_{H} (\pi^N * \iota_\epsilon) 
\; +\; \bb V_{N,\epsilon}^{H} \;+\; C_{H} (\epsilon)\Big\} \; 
\; ,
$$
where $\lim_{\epsilon \to 0} C_{H} (\epsilon)=0$, $\iota_\epsilon$
stands for the approximation of the identity $\iota_\epsilon (u) = (2
\epsilon)^{-1} \mb 1 \{ u\in [- \epsilon, \epsilon]\}$, $*$ stands for
convolution,
$$
\bb V_{N,\epsilon}^{H} \;=\;  \int_0^{T}
V_{N,\epsilon}^{H,\Psi_0}(t,\eta_t)\, dt  
\; +\;  W_{ \nabla H(\cdot, 1)}^+ \; 
-\; W_{ \nabla H(\cdot, -1)}^- 
$$
and $\Psi_0 (\eta) = \eta(0) [1-\eta(1)]$.

Fix a subset $A$ of $D([0,T],\mc M)$ and write 
$$
\frac 1N \log \bb P_{\eta^N} [\pi^N\in A ] \;=\; 
\frac 1N \log \bb E_{\eta^N} \Big[ M^H_T (M^H_T)^{-1}
\mb 1\{\pi^N \in A\} \Big]\; .
$$
Maximizing over $\pi^N$ in $A$, we get from previous computation
that the last term is bounded above by
$$
- \inf _{\pi\in A} J_{H} (\pi * \iota_\epsilon)  
\;+\; \frac 1N \log \bb E_{\eta^N} \Big[    
M^{H}_T e^{- N  \bb V_{N,\epsilon}^{H}} \Big] \;-\;
C_{H} (\epsilon) \; .
$$
Denote by $\bb P_{\eta^N}^{H}$ the measure $\bb P_{\eta^N}
M^{H}_T$. Since the martingale is bounded by $\exp\{ CN\}$ for some
finite constant depending only on $H$ and $T$, Theorem \ref{s04} holds
for $\bb P_{\eta^N}^{H}$ in place of $\bb P_{\eta^N}$.  In particular,
the second term of the previous formula is bounded above by
$C_H(\epsilon, N)$ such that $\lim_{\epsilon\to 0}
\limsup_{N\to\infty} C_H(\epsilon, N) =0$.  Hence, for every
$\epsilon>0$, and every $H$ in $C^{1,2}_0([0,T]\times [-1,1])$,
$$
\limsup_{N\to\infty} \frac 1N \log \bb P_{\eta^N} [\pi^N\in A ]
\;\le\;  - \inf _{\pi\in A} J_{H} (\pi * \iota_\epsilon)  
\; +\;  C_H'(\epsilon)\; ,
$$
where $\lim_{\epsilon\to 0} C_H'(\epsilon) =0$. 

Assume now that the set $A$ is a compact set $K$. Since $J_H (\cdot *
\iota_\epsilon)$ is continuous for every $H$ and $\epsilon >0$, we may
apply the arguments presented in Lemma 11.3 of \cite{v} to exchange
the supremum with the infimum.  In this way we obtain that the last
expression is bounded above by
$$
\limsup_{N\to\infty} \frac 1N \log \bb P_{\eta^N} [\pi^N\in K ]
\;\le\; - \inf _{\pi\in K} \sup_{H,\epsilon} 
\Big\{ J_{H} (\pi * \iota_\epsilon) \; +\;  C_H'(\epsilon) \Big\} \; .
$$
Letting first $\epsilon\downarrow 0$, since $J_H(\pi *
\iota_\epsilon)$ converges to $J_H(\pi)$ for every $H$ in
$C^{1,2}_0([0,T]\times [-1,1])$, in view of the definition
(\ref{eq:3}) of $I_T(\pi | \gamma)$, we deduce that 
$$
\limsup_{N\to\infty} \frac 1N \log \bb P_{\eta^N} [\pi^N\in K ]
\;\le\; - \inf _{\pi\in K} I_T (\pi | \gamma) \; ,
$$
which proves the upper bound for compact subsets.

To pass from compact sets to closed sets, we have to obtain
``exponential tightness'' for the sequence $\bb P_{\eta^N}[\pi^N \in
\cdot]$. The proof presented in \cite{b} for the non interacting
zero-range process is easily adapted to our context.

\subsection{Hydrodynamic limit of weakly asymmetric exclusions}
\label{s:3.4}

Fix a function $H$ in $C^{1,2}_0([0,T]\times [-1,1])$ and recall the
definition of the martingale $M_T^H$. Denote by $\bb P_{\eta^N}^H$ the
probability measure on $D([0,T], \Sigma_N)$ defined by $\bb
P_{\eta^N}^H [A] = \bb E_{\eta^N} [ M_T^H \mb 1\{A\}]$. Under $\bb
P_{\eta^N}^H$, the coordinates $\{\eta_t :\, 0\le t\le T\}$ form a
Markov process with generator $L_N^H = L_{+,N} + L_{0,N}^H + L_{-,N}$,
where 
$$
(L_{0,N}^H f)(\eta) \;=\; \frac {N^2}2 \sum_{x=-N}^{N-1} e^{-
  \{H(t,[x+1]/N) - H(t,x/N)\} \{\eta(x+1) - \eta (x)\}} [f(\sigma
^{x,x+1}\eta) - f(\eta)]\; .
$$

The next result is due to Eyink, Lebowitz and Spohn \cite{els2}.
Recall $\chi(\rho)=\rho(1-\rho)$.

\begin{lemma}
\label{s05}
Consider a sequence $\eta^N$ associated to some profile 
$\gamma \in\mc M$ and fix $H$ in $C^{1,2}_0([0,T]\times [-1,1])$. Then,
for all $t>0$, $\pi^N(t) = \pi^N(\eta_{t})$ converges in probability
(in the sense \reff{f01}) to $\rho(t,u)$, the unique weak solution of
\begin{equation}
\label{f06}
\left\{
  \begin{array}{l}
\partial_t \rho = (1/2) \Delta \rho -  \nabla \{ \chi(\rho)
 \nabla H \}\; , \\
\rho(t, \pm 1) = \rho_\pm \; , \\
\rho(0, \cdot) = \gamma (\cdot)\; .
  \end{array}
\right.
\end{equation}
\end{lemma}

As in subsection \ref{s:3.1}, by a weak solution of the Dirichlet problem
\reff{f06} in the time interval $[0,T]$, we understand a {\it bounded}
real function $\rho$ which satisfies the following two conditions.
\begin{description}
\item [(a)] There exists a function $A(t,u)$ in $L^2([-1,1]\times
[0,T])$ such that 
\begin{eqnarray*}
&& \int_0^t ds \int_{-1}^1 du \, \rho(s,u) ( \nabla G)(u) \\
&& \qquad\qquad\quad
=\; \{\rho_+ G(1) - \rho_- G(-1) \} t
\; -\;  \int_0^t ds \int_{-1}^1 du \, A(s,u) G(u) 
\end{eqnarray*}
for every smooth function $G\colon [-1,1]\to \bb R$ and every $0\le
t\le T$. $A(t,u)$ will be denoted by $( \nabla \rho)(t,u)$.

\item[(b)] For every function $G\in C^1\big([-1,1]\big)$ vanishing at
  the boundary and every $t\ge 0$,
\begin{eqnarray*}
&& \int_{-1}^1 du \, \rho(t,u) G(u) \; -\; 
\int_{-1}^1 du \, \gamma(u) G(u) \; = \\
&& \quad
\int_0^t ds \int_{-1}^1 du \, ( \nabla G)(u)
\Big\{ - (1/2) ( \nabla \rho)(s,u) + 
\chi( \rho(s,u)) ( \nabla H)(s,u) \Big\} \; .
\end{eqnarray*}
\end{description}
 
The classical $H_{-1}$ estimates gives uniqueness of weak solutions of
equation \reff{f06}.

\subsection{The rate function}

We prove in this subsection some properties of the rate function $I_T
(\,\cdot\, | \gamma)$. We first claim that this rate function is
convex and lower semi continuous. In view of the definition of $I_T
(\,\cdot\, | \gamma)$, to prove this assertion, it is enough to show
that $J_H$ is convex and lower semi continuous for each $H$ in
$C^{1,2}_0([0,T]\times [-1,1])$.  It is convex because $\chi(a) = a(1-a)$
is a concave function. It is lower semi continuous because for any
positive, continuous function $G:[0,T]\times [-1,1]\to \bb R$ and for
any sequence $\pi^n$ converging to $\pi$ in $D([0,T], \mc M)$,
\begin{eqnarray*}
\int_0^T dt\, \langle\chi(\pi(t)) , G(t) \rangle &=&
\lim_{\epsilon \to 0} \int_0^T dt\, \langle\chi(\pi(t) * \iota_\epsilon) , 
G(t)\rangle \\
&=&  \lim_{\epsilon \to 0} \lim_{n\to\infty} \int_0^T dt\, 
\langle\chi(\pi^n(t) * \iota_\epsilon) , G(t)\rangle\; .
\end{eqnarray*}
Since $\chi$ is concave and $G$ positive, a change of variables shows
that this expression is bounded below by
$$
\lim_{\epsilon \to 0} \limsup_{n\to\infty} \int_0^T dt\, 
\langle\chi(\pi^n(t) ) , G(t) * \iota_\epsilon\rangle\; =\;
\limsup_{n\to\infty} \int_0^T dt\,  \langle\chi(\pi^n(t) ) , G(t) \rangle
$$
because $G$ is continuous and $\chi$ is bounded. This proves that $J_H$
is lower semi continuous for every $H$ in $C^{1,2}_0([0,T]\times
[-1,1])$.

Denote by $D_\gamma$ the subset of $D([0,T], \mc M)$ of
all paths $\pi(t,u)$ which satisfy the boundary conditions $\pi(0,
\cdot) = \gamma (\cdot)$, $\pi(\cdot, \pm 1) = \rho_\pm$,
in the sense that for every $0\le t_0 <t_1\le T$,
$$
\lim_{\delta\to 0} \int_{t_0}^{t_1} dt\, \frac 1\delta 
\int_{-1}^{-1+\delta} \pi(t,u) \, du \; =\; (t_1-t_0) \rho_- 
$$
and a similar identity at the other boundary.

\begin{lemma}
\label{s06}
$I_T (\pi | \gamma) \;=\; \infty$ if $\pi$ does not belong to $D_\gamma$.
\end{lemma}

\begin{demo}
  Fix $\pi$ in $D([0,T], \mc M)$ such that $I_T (\pi | \gamma) <\infty$.
  We first show that $\pi (0, \cdot ) = \gamma (\cdot)$. For
  $\delta>0$, consider the function $H_\delta (t,u) = h_\delta (t)
  g(u)$, $h_\delta (t) = (1- \delta^{-1}t)^+$, $g(\cdot)$ vanishing at
  the boundary $\pm 1$. Here $a^+$ stands for the positive part of
  $a$. Of course, $H_\delta$ can be approximated by smooth functions.
  Since $\pi$ is bounded and since $t \to \pi(t, \cdot)$ is right
  continuous for the weak topology, 
$$
\lim_{\delta \downarrow 0} J_{H_\delta}(\pi) \;=\; \langle\pi(0), g\rangle - \langle\gamma,
g\rangle\; ,
$$
which proves that $\pi(0) = \gamma$ a.s.\ because $I_T (\pi | \gamma)
<\infty$.

A similar argument shows that $\pi(t,\pm 1) = \rho_\pm$; to prove this
statement we may consider the sequence of functions $H_\delta (t,u) =
h(t) g_\delta (u)$, where $h(t)$ approximates the indicator of some
time interval $[t_0, t_1]$ and where
$$
g_\delta' (u) \;=\;
\left\{ 
  \begin{array}{ll}
\vphantom{\Big\{}
A - (A+b) (1 + u) / \delta & \text{if $-1 \le u\le
  -1+\delta$, } \\
\vphantom{\Big\{}
-b & \text{if $-1 + \delta \le u\le 1$. }
  \end{array}
\right.
$$
Here $A>0$ is large and fixed and $b=b(A,\delta)>0$ is chosen for
the integral over $[-1,1]$ of $g_\delta'$ to vanish.
\end{demo}

Fix $\pi$ in $D_\gamma$ and denote by $\mc H_1(\pi)$ the Hilbert
space induced by  $C^{1,2}_0([0,T]\times [-1,1])$ endowed with the
inner product $\langle\cdot , \cdot\rangle_\pi$ defined by
$$
\langle H,G\rangle_\pi \; =\; \int_0^T dt \int_{-1}^1 du\, \chi(\pi) ( \nabla G)
( \nabla H)\; .
$$

\begin{lemma}
\label{s07}
Fix a trajectory $\pi$ in $D_\gamma$ and assume that $I_T (\pi |
\gamma)$ is finite.  There exists a function $H$ in $\mc H_1(\pi)$
such that $\pi$ is the unique weak solution of
\begin{equation}
\label{f07}
\left\{
  \begin{array}{l}
\vphantom{\Big\{}
\partial_t \pi = (1/2) \Delta \pi -  \nabla \{ \chi(\pi)
\pi (1-\pi)
 \nabla H \}\; , \\
\vphantom{\Big\{}
\pi(t, \pm 1) = \rho_\pm \; , \\
\vphantom{\Big\{}
\pi(0, \cdot) = \gamma (\cdot)\; .
  \end{array}
\right.
\end{equation}
Moreover, 
\begin{equation}
\label{f03}
I_T (\pi | \gamma) \; =\; (1/2) \int_0^T dt \int_{-1}^1 du\, \chi(\pi)
( \nabla H)^2\; .
\end{equation}
\end{lemma}

We refer the reader to \cite{kl,kov} for a proof.  One of the
consequences of this lemma is that every trajectory $t\mapsto \pi(t)$
with finite rate function is continuous in the weak topology,
$\pi\in C([0,T];\mc M)$.  Indeed,
by the previous lemma, for $\pi$ such that $I_T (\pi | \gamma)
<\infty$, and every $G$ in $C^2_0\big([-1,1]\big)$,
\begin{eqnarray*}
\langle\pi(t), G\rangle - \langle\pi(s), G\rangle &=& 
(1/2) \int_s^t dr\, \langle\pi(r), \Delta G\rangle
+ \int_s^t dr\, \langle\chi(\pi(r) ),  \nabla G \,  \nabla H\rangle \\
&& 
- (1/2)\{ ( \nabla G)(1) \, \rho_+ - ( \nabla G)(-1)
\, \rho_-\}(t-s)
\end{eqnarray*}
for some $H$  in $\mc H_1(\pi)$. Since $G$ is smooth and $H$ belongs
to $\mc H_1(\pi)$, the right hand side vanishes as $|t-s|\to 0$.

\subsection{Lower bound}
\label{s:lb}
Denote by $D^0_\gamma$ the set of trajectories $\pi$ in $D([0,T], \mc
M)$ for which there exists $H$ in $C^{1,2}_0([0,T]\times [-1,1])$ such
that $\pi$ is the solution of \reff{f07}. For each $\pi$ in
$D^0_\gamma$, and for each neighborhood $\mc N_\pi$ of $\pi$
$$
\liminf_{N\to\infty} \frac 1N \log \bb P_{\eta^N} [ \pi^N \in \mc N_\pi ]
\;\ge\; - I_T (\pi | \gamma) \; .
$$

This statement is proved as in the periodic boundary case, see
\cite{kl}.  To complete the proof of the lower bound, it remains to
show that for every trajectory $\pi$ such that $I_T (\pi | \gamma)
<\infty$, there exists a sequence $\pi_k$ in $D^0_\gamma$ such that
$\lim_k \pi_k = \pi$, $\lim_k I_T (\pi_k | \gamma) = I_T (\pi |
\gamma)$.

This is not too difficult in our context because the rate function is
convex and lower semi continuous.  We first show that any path $\pi$
with finite rate function can be approximated by a path which is
bounded away from $0$ and $1$.  Fix a path $\pi$ such that $I_T (\pi |
\gamma) <\infty$. Fix $\delta > 0$ and denote by $\rho(t,u)$ the
solution of the hydrodynamic equation \reff{f02} with initial
condition $\gamma$ instead of $\rho_0$. Let $\pi_\delta = \delta
\rho + (1-\delta) \pi$. Of course, $\pi_\delta$ converges to $\pi$
as $\delta\downarrow 0$. By lower semi continuity, $I_T (\pi | \gamma)
\le \liminf_{\delta\to 0} I_T (\pi_\delta | \gamma)$. On the other
hand, since $I_T (\,\cdot\, | \gamma)$ is convex, $I_T (\pi_\delta |
\gamma) \le (1-\delta) I_T (\pi | \gamma)$ because $\rho$ is the
solution of the hydrodynamic equation and $I_T (\rho| \gamma) = 0$.
This shows that $\lim_{\delta\to 0} \pi_\delta = \pi$,
$\lim_{\delta\to 0} I_T (\pi_\delta | \gamma) = I(\pi)$. Since
$0<\gamma <1$, $0<\rho_\pm <1$, $\pi_\delta$ is bounded away from
$0$ and $1$, proving the claim.

Fix now a path $\pi$ with finite rate function and bounded away from
$0$ and $1$. We claim that this trajectory may be approximated by a
path in $D_\gamma^0$. Since $I_T (\pi | \gamma) <\infty$, by Lemma
\ref{s07}, there exists $H$ in $\mc H_1(\pi)$ satisfying \reff{f07}.
Since $\pi$ is bounded away from $0$ and $1$, $\mc H_1(\pi)$ coincides
with the usual Sobolev space $H_1$ associated to the Lebesgue measure.
Consider a sequence of smooth functions $H_n : [0,T]\times [-1,1]\to
\bb R$ vanishing at the boundary and such that $ \nabla H_n$
converges in $L^2([0,T]\times [-1,1])$ to $ \nabla H$.  Denote by
$\pi^n$ the solution of \reff{f06} with $H_n$ instead of $H$. We claim
that $\lim_{n\to\infty} \pi^n = \pi$, $\lim_{n\to\infty} I_T (\pi^n |
\gamma) = I_T (\pi | \gamma)$.

The proof that $\pi^n$ converges to $\pi$ is divided in two pieces. We
first show that the sequence is tight in $C([0,T], \mc M)$ and then we
prove that all limit points are solution of equation \reff{f06}. We
start with a preliminary estimate which will be needed
repeatedly. Recall that $\bar \rho$ is the stationary
profile. Computing the time derivative of $\int_{-1}^1 du (\pi^n(t) -
\bar \rho)^2$, we obtain that
\begin{equation}
\label{f10}
\int_0^T dt\, \int_{-1}^1 du\, ( \nabla \pi^n(t) )^2 \;\le\; C
\end{equation}
for some finite constant independent of $n$. 

 From the previous bound and since $\pi^n(t,u)$ belongs to $[0,1]$, it
is not difficult to show that the sequence $\pi^n$ is tight in
$C([0,T], \mc M)$. To check uniqueness of limit points, consider any
limit point $\beta$ in $C([0,T], \mc M)$. We claim that $\beta$ is a
weak solution of the equation \reff{f06}. Of course $\beta$ is
positive and bounded above by $1$.  The existence of a function
$A(s,u)$ in $L^2([-1,1]\times [0,T])$ for which {\bf (a)} holds
follows from \reff{f10}, which guarantees the existence of weak
converging subsequences. The unique difficulty in the proof of
identity {\bf (b)} is to show that for any $0\le t\le T$, $G$ in
$L^2([0,T]\times [-1,1])$,
\begin{equation}
\label{f11}
\lim_{n\to\infty} \int_0^t ds\, \langle\chi(\pi^n(s)), G(s)\rangle \; =\; 
\int_0^t ds\, \langle\chi(\beta(s)), G(s)\rangle
\end{equation}
for any sequence $\pi^n$ converging to $\beta$ in $C([0,T], \mc M)$ and
satisfying \reff{f10}.  This identity holds because for any $\delta >0$
$$
\lim_{n\to\infty} \int_0^t ds\, \langle\chi(\pi^n(s)*\iota_\delta), G(s)\rangle \; 
=\; \int_0^t ds\, \langle \chi (\beta(s)*\iota_\delta), G(s)\rangle
$$
and because, by Schwartz inequality and $|\chi(a)-\chi(b)|\le |a-b|$,
$$
\begin{array}{l}
{\displaystyle
\Big( \int_0^t ds\, \langle \chi(\pi^n(s)*\iota_\delta) - 
\chi (\pi^n(s)), G(s)\rangle\Big)^2
}
\\
\quad\quad
{\displaystyle
\le\;
\int_0^t ds\, \langle G(s)^2\rangle \int_0^t ds\, 
\langle [\pi^n(s)*\iota_\delta - \pi^n(s)]^2\rangle\; .
}
\end{array}
$$
It is not difficult to show, using estimate \reff{f10}, that this term
vanishes as $\delta\downarrow 0$, uniformly in $n$, proving
\reff{f11}. In conclusion, we proved that the sequence $\pi^n$ is
tight in $C([0,T], \mc M)$ and that all its limit points are weak
solutions of equation \reff{f06}. By uniqueness of weak solutions,
$\pi^n$ converges in $C([0,T], \mc M)$ to $\pi$.

It remains to see that $I_T (\pi^n | \gamma)$ converges to $I_T (\pi |
\gamma)$. Since $\pi^n \to \pi$ and $I_T (\,\cdot\, | \gamma)$ is
lower semi continuous, we just need to check that $\limsup_n I_T
(\pi^n | \gamma) \le I_T (\pi | \gamma)$. Here again the concavity and
the boundness of $\chi$ help. Since $ \nabla H^n$ converges in $L^2$
to $ \nabla H$ and $\chi$ is bounded, the main problem is to show that
$$
\limsup_{n\to\infty} \int_0^T dt\, \langle\chi(\pi^n(t)) , ( \nabla H(t))^2\rangle
\;\le\; \int_0^T dt\, \langle\chi(\pi(t)) , ( \nabla H(t))^2\rangle\; .
$$
Since $\pi*\iota_\delta$ converges almost surely to $\pi$ as
$\delta\downarrow 0$,
\begin{eqnarray*}
\int_0^T dt\, \langle\chi(\pi(t)) , ( \nabla H(t))^2\rangle &=&
\lim_{\delta\to 0} \int_0^T dt\, \langle\chi(\pi(t)*\iota_\delta) , 
( \nabla H(t))^2\rangle \\
&=& \lim_{\delta\to 0} 
\lim_{n\to\infty} \int_0^T dt\, \langle\chi(\pi^n(t) *\iota_\delta) , 
( \nabla H(t))^2\rangle\; .
\end{eqnarray*}
Since $\chi$ is concave, the previous expression is bounded below by
$$
\lim_{\delta\to 0} 
\limsup_{n\to\infty} \int_0^T dt\, \langle\chi(\pi^n(t))*\iota_\delta , 
( \nabla H(t))^2\rangle\; .
$$
Since $\chi$ is bounded and $( \nabla H)^2$ integrable, a change of
variables shows that the previous expression is equal to
$$
\limsup_{n\to\infty} \int_0^T dt\, \langle\chi(\pi^n(t)) , 
( \nabla H(t))^2\rangle\; ,
$$
concluding the proof of the lower bound.

\section{The rate function for the invariant measure}
\label{s:4}
In this section we discuss some properties of the functional $S(\rho)$
which are needed later.  The results stated here are essentially
contained in \cite{DLS}, but, for the sake of completeness, we review
them and give more detailed proofs.  Without any loss of generality,
from now on we shall assume that $0 < \rho_- < \rho_+ < 1$.
Recall the definitions of the set $\mc F$, \reff{dF},
and of the functional $\mc G (\rho, f)$, \reff{dG}.  

The Euler--Lagrange equation associated to the variational problem
\reff{SsS} is given by the non linear boundary value problem
\begin{equation}\label{Deq_old}
\left\{
\begin{array}{l}
\vphantom{\Big(}
{\displaystyle F'' = 
\big( \rho - F \big) \frac{\big( F' \big)^2}{F(1-F)} 
\qquad \textrm{ in $(-1,1)$}  }\; ,
\\
\vphantom{\Big(}
{\displaystyle F(\pm 1) = \rho_\pm }\; .
\end{array}
\right.
\end{equation}
We introduce the notation, which we will use throughout this 
section,
\begin{equation}\label{referee}
\rr(u)=\rr(\rho,F;u)=
(\rho(u)-F(u))\frac{F^\prime(u)}{F(u)(1-F(u))}\ .
\end{equation}
Using this notation equation (\ref{Deq_old}) takes the form
\begin{equation}\label{Deq}
\left\{
\begin{array}{l}
\vphantom{\Big(}
{\displaystyle F'' = 
F^\prime\rr
\qquad \textrm{ in $(-1,1)$}  }\; ,
\\
\vphantom{\Big(}
{\displaystyle F(\pm 1) = \rho_\pm }\; .
\end{array}
\right.
\end{equation}

In order to state and prove an existence and uniqueness result for
$F\in\mc F$ we formulate \reff{Deq} as the integro--differential
equation 
\begin{equation}\label{DeqI}
F(u) = \rho_- + ( \rho_+- \rho_-) 
\frac{ 
\displaystyle
\int_{-1}^u \!dv \, 
\exp\left\{ \int_{-1}^v\!dw \, 
\rr(\rho,F;w)
\right\}}
{\displaystyle
\int_{-1}^1 \!dv \, 
\exp\left\{ \int_{-1}^v\!dw \, 
\rr(\rho,F;w)
\right\}}\; .
\end{equation}
We will denote its solution by $F= F(\rho)$ to emphasize its
dependence on $\rho$. We observe that if $\rho=\bar\rho$ then $F=F
(\bar\rho) = \bar\rho$ solves \reff{DeqI} and \reff{Deq}.

Notice that if $F\in C^2\big([-1,1]\big)$ is a solution of the
boundary value problem (\ref{Deq}) such that $F^\prime(u)>0$ for
$u\in[-1,1]$, then $F$ is also a solution of the integro--differential
equation (\ref{DeqI}).  Conversely, if $F\in C^1\big([-1,1]\big)$ is a
solution of (\ref{DeqI}), then $F^\prime(u)>0$, $F^{\prime\prime}(u)$
exists for almost every $u$ and (\ref{Deq}) holds almost everywhere.
Moreover, if $\rho\in C\big([-1,1]\big)$, then $F\in
C^2\big([-1,1]\big)$ and (\ref{Deq}) holds everywhere.

\begin{remark}\label{rem42}
There are non monotone solutions of equation (\ref{Deq}).  For
example, for the constant profile $\rho=1/2$, it is easy to check that
the functions
$$
F(u) = \frac 12 \left[ 1 + \sin\big( \lambda u + \varphi \big) \right]
\; ;
$$
satisfy equation (\ref{Deq}) for countably many choices of the
parameters $\lambda$ and $\varphi$ (fixed in order to satisfy the
boundary conditions in (\ref{Deq})).  However only one such function
is monotone.  In fact, under the monotonicity assumption on $F$, we
will prove uniqueness (and existence) of the solution of the boundary
value problem (\ref{Deq}).
\end{remark}

The following theorem gives us the existence and uniqueness result for
\reff{DeqI} together with a continuous dependence of the solution on
$\rho$. Recall that we denote by $C^1\big([-1,1]\big)$ the Banach
space of continuously differentiable functions $f :[-1,1]\to \bb
R$ endowed with the norm $\|f\|_{C^1} := \sup_{u\in[-1,1]} \big\{
|f(u)| + |f'(u)| \big\}$.

\begin{theorem}
\label{t:deq} 
For each $\rho\in\mc M$ there exists in $\mc F$ a unique solution 
$F=F(\rho)$ of \reff{DeqI}. Moreover:
$\phantom{merd}$\par\noindent
\begin{itemize}
\item[{(i)}]{if $\rho\in C\big([-1,1]\big)$, then 
$F=F(\rho)\in C^2\big([-1,1]\big)$ and 
it is the unique solution in $\mc F \cap C^2\big([-1,1]\big)$ of
\reff{Deq};}
\item[{(ii)}]{if $\rho_n$ converges to $\rho$ in $\mc M$ as
$n\to\infty$, then $F_n=F(\rho_n)$ converges to 
$F=F(\rho)$ in $C^1\big([-1,1]\big)$;}
\item[{(iii)}]{fix $T>0$ and consider a function 
$\rho=\rho(t,u)\in C^{1,0}\big([0,T]\times[-1,1]\big)$. 
Then $F=F(t,u)= F(\rho(t,\cdot))(u)\in 
C^{1,2}\big([0,T]\times[-1,1]\big)$.} 
\end{itemize}
\end{theorem}


The existence result in Theorem \ref{t:deq} will be proven by applying
Schauder's fixed point theorem.  
For each $\rho\in\mc{M}$ consider the map $\mc K_\rho : \mc{F} \to
C^1\big([-1,1]\big)$ given by
\begin{equation}\label{Kg} 
\mc K_\rho(f) (u) :=  
\rho_- + ( \rho_+- \rho_-) 
\frac{ 
\displaystyle
\int_{-1}^u \!dv \, 
\exp\left\{ \int_{-1}^v\!dw \, 
\rr(\rho,f;w)
\right\}}
{\displaystyle
\int_{-1}^1 \!dv \, 
\exp\left\{ \int_{-1}^v\!dw \, 
\rr(\rho,f;w)
\right\}}\; \cdot
\end{equation}
Let us also define the following closed, convex subset 
of $C^1\big([-1,1]\big)$:
\begin{equation} 
\label{v2}
\mc B := \left\{ f \in C^1\big([-1,1]\big) \,:\: f(\pm 1)
=\rho_\pm,\: b\le f'(u) \le B  \, 
\right\}\subset\mc{F}\; ,
\end{equation}
where, recalling we are assuming $\gamma_-<\gamma_+$,
$$
b := \frac{ \rho_+- \rho_-}{2} \, 
\frac{\gamma_-}{\gamma_+}
\ , \quad
B := \frac{ \rho_+- \rho_-}{2} \, 
\frac{\gamma_+}{\gamma_-}\ .
$$

\begin{lemma}
\label{t:sch}
For each $\rho\in\mc M$, $\mc K_\rho$ is a continuous map on $\mc F$
and $\mc K_\rho(\mc F)\subset\mc B$.
Furthermore
$\mc{K}_\rho(\mc{B})$ has compact closure in $C^1\big([-1,1]\big)$.
Hence, by Schauder's fixed point theorem, for each $\rho\in\mc M$
equation \reff{DeqI} has a solution $F= \mc K_\rho(F)\in \mc{B}$.
Moreover, 
there exist a constant
$C\in(0,\infty)$ depending on $\rho_\pm$ such that for any
$\rho\in\mc{M}$ and any $u,v\in [-1,1]$ we have 
$|F'(u)-F'(v)|\le C \,|u-v|$.
\end{lemma}
 
\begin{demo}
It is easy to check that $\mc K_\rho $ is continuous and $\mc K_\rho
(f) (\pm 1) = \rho_\pm$. Let us define $g_\rho:= \mc K_\rho (f)$, we
have
\begin{equation}\label{g'}
g_\rho'(u) = ( \rho_+- \rho_-) 
\frac{ 
\displaystyle
\exp\left\{ \int_{-1}^u\!dw \, 
\rr(\rho,f;w)
\right\}}
{\displaystyle
\int_{-1}^1 \!dv \, 
\exp\left\{ \int_{-1}^v\!dw \, 
\rr(\rho,f;w)
\right\}}\; \cdot
\end{equation}
Since $\rho(w) - f(w)\le 1 - f(w)$, $\rho(w) - f(w)\ge  - f(w)$, and
$f'(w) \ge 0$, we get
$$
\frac{(1-f)^\prime}{1-f}\leq\rr\leq\frac{f^\prime}{f}
$$
which implies $b\leq g_\rho'(u)\leq B$ for all $u\in[-1,1]$.
In particular
$\mc K_\rho (\mc F) \subset \mc B$. 

To show that $\mc K_\rho (\mc B)$ has a compact closure, by Ascoli--Arzela
theorem, it is enough to prove that $g_\rho'$ is Lipschitz uniformly
for $f\in\mc B$. Indeed, by using \reff{g'}, it is easy to check
that there exists a constant $C=C(\rho_-,\rho_+)<\infty$ 
such that for any $u,v\in [-1,1]$, any $f\in \mc B$, and any
$\rho\in\mc M$ we have
$\big| g_\rho'(u)-g_\rho'(v) \big| \le C |u-v|$.
\end{demo}

\noindent{{\bf Proof of Theorem \ref{t:deq}:}~} 
The existence of solutions for \reff{DeqI} has been proven in Lemma
\ref{t:sch}; to prove uniqueness we follow closely the argument in
\cite{DLS}. Consider a solution $F\in\mc{F}$ of (\ref{DeqI}).
Since it solves (\ref{Deq}) almost everywhere, we get
\begin{equation}
\label{u1.5}
F'(u) \; =\; F'(-1) \; +\; \int_{-1}^u\!dw
\, 
F^\prime(w)\rr(\rho,F;w)
\end{equation}
for all $u$ in $[-1,1]$. Moreover, taking into account that $F$ is
strictly increasing, we get from (\ref{Deq}) that
$$
\Big( \frac {F(1-F)}{F'} \Big)' \; =\; 1 - F - \rho
$$
holds a.e., so that
\begin{equation}
\label{u2}
\frac{ F(u) [ 1-F(u)] } {F'(u)}
=\frac{\rho_- [ 1-\rho_-] } {F'(-1)}
+ \int_{-1}^u \!dv \, \left[ 1-F(v) -\rho(v) \right]
\end{equation}
for all $u$ in $[-1,1]$.

Let $F_1,F_2\in \mc F$ be two solutions of \reff{DeqI}.  If
$F_1'(-1)=F_2'(-1)$ an application of Gronwall inequality in
(\ref{u1.5}) yields $F_1=F_2$.  We next assume $F_1'(-1)< F_2'(-1)$
and deduce a contradiction.  Keep in mind that $F_i'>0$ because $F_i$
belongs to $\mc F$ and recall (\ref{u2}).  Let $\bar u := \inf\{ v\in
(-1,1] \,:\: F_1(v) = F_2(v) \}$ which belongs to $(-1,1]$ because
$F_1(\pm 1)= F_2(\pm 1 )$ and $F_1'(-1)<F_2'(-1)$. By definition of
$\bar u$, $F_1(u) < F_2(u)$ for any $u\in (-1,\bar u)$,
$F_1(\bar{u})=F_2(\bar{u})$ and $F_1^\prime(\bar{u})\geq F_2^\prime(\bar{u})$.
By \reff{u2},
we also obtain 
$$
\frac{ F_1(\bar u) [ 1-F_1(\bar u)] } {F_1'(\bar u)} > 
\frac{ F_2(\bar u) [ 1-F_2(\bar u)] } {F_2'(\bar u)}
$$
or, equivalently, $F_1'(\bar u) < F_2'(\bar u)$, which is a
contradiction and concludes the proof of the first statement of
Theorem \ref{t:deq}.
\smallskip

We turn now to statement (i). Existence follows from identity
(\ref{Deq}), which now holds for all points $u$ in $[-1,1]$ because
$\rho$ is continuous. Uniqueness follows from the uniqueness for the
integro--differential formulation \reff{DeqI}.
\smallskip

To prove ({ii}), let $\rho_n$ be a sequence converging to $\rho$
in $\mc M$ and denote by $F_n=F(\rho_n)$ the corresponding solution of
\reff{DeqI}. By Lemma \ref{t:sch} and Ascoli--Arzela theorem, the
sequence $F_n$ is relatively compact in $C^1\big([-1,1]\big)$. It
remains to show uniqueness of its limit points.  
Consider a subsequence $n_j$ and assume that $F_{n_j}$ converges to
$G$ in $C^1\big( [-1,1]\big)$. Since $\rho_{n_j}$ converges to $\rho$
in $\mc M$ and $F_{n_j}$ converges to $G$ in $C^1\big( [-1,1]\big)$,
by \reff{Kg} $\mc K_{\rho_{n_j}} (F_{n_j})$ converges to $\mc K_{\rho}
(G)$. In particular, $G = \lim_j F_{n_j} = \lim_j \mc K_{\rho_{n_j}}
(F_{n_j}) = \mc K_{\rho} (G)$ so that, by the uniqueness result,
$G=F(\rho)$.  This shows that $F(\rho)$ is the unique possible limit
point of the sequence $F_n$, and concludes the proof of (ii).

\smallskip
We are left to prove (iii). 
If $\rho(t,u)\in C^{1,0}\big([0,T]\times[-1,1]\big)$,
we have from (i) and (ii) that
$F(t,u)=F(\rho(t,\cdot))(u)\in C^{0,2}\big([0,T]\times[-1,1]\big)$.
We then just need to prove that $F(t,u)$, as a function of $t$,
is continuously differentiable.
This will be accomplished by Lemma \ref{t:ddeq} below.
{\qed \medskip}

In order to prove the differentiability of $t\mapsto F(t,u):=
F(\rho(t,\cdot))(u) $ it is convenient to introduce the new variable
\begin{equation}
\f (t,u) := \log \frac {F(t,u)}{1-F(t,u)},
\qquad (t,u)\in [0,T]\times [-1,1]
\end{equation}
Note that $\f \in [\f_-,\f_+]$ where 
$\f_\pm := \log[ \rho_\pm / (1-\rho_\pm)] = \log \gamma_\pm$ and 
$u\mapsto \f(t,u)$ is strictly increasing.
We remark that, as discussed in \cite{BDGJL}, while the function $F$ is
analogous to a density, the variable $\f$ can be interpreted as a
thermodynamic force.
The advantage of using $\f$ instead of $F$ lies in the fact that, as a
function of $\f$, the functional $\mc G$ is concave. This property
plays a crucial role in the sequel.

Let us fix a density profile 
$\rho\in C^{1,0}\left([0,T]\times [-1,1] \right)$. By 
(i)--(ii) in Theorem \ref{t:deq} and elementary
computations, we have that $\f\in C^{0,2}\left([0,T]\times [-1,1]
\right)$ and it is the unique strictly increasing (w.r.t.\ $u$) 
solution of the problem 
\begin{equation}\label{Deqf}
\left\{
\begin{array}{ll}
{\displaystyle \vphantom{\Big(}
\frac{\Delta \f (t,u)}{ \left(  \nabla \f (t,u) \right)^2}
+ \frac{1}{ 1+ e^{\f(t,u)}} = \rho(t,u) 
} 
&{\displaystyle  (t,u) \in [0,T]\times(-1,1) }
\\
{\displaystyle 
\vphantom{\Big(}
\f(t,\pm 1) = \f_\pm 
}
&
{\displaystyle 
t\in [0,T]
}
\end{array}
\right.
\end{equation}
Note also that, by Lemma \ref{t:sch}, there exists a constant
$C_1=C_1(\rho_-,\rho_+)\in (0,\infty)$ such that
\begin{equation}\label{sf'}
\frac{1}{C_1} \le  \nabla \f (t,u) \le C_1
\qquad \forall \, (t,u) \in [0,T]\times[-1,1] 
\end{equation}

\begin{lemma}
\label{t:ddeq}
Let $\rho\in C^{1,0}\left([0,T]\times [-1,1] \right)$ 
and $\f=\f(t,u)$ be the corresponding solution of \reff{Deqf}.
Then $\f \in C^{1,2}\left([0,T]\times [-1,1] \right)$ and 
$\psi(t,u):= \partial_t \f(t,u)$ is the unique classical solution of
the linear boundary value problem
\begin{equation}\label{dDeqf}
 \nabla\Big[ 
\frac{ \nabla \psi (t,u)}{ \big(  \nabla \f (t,u) \big)^2}
\Big]
- \frac{e^{\f(t,u)}}{ \big( 1+ e^{\f(t,u)}\big)^2} \, \psi(t,u) 
= \partial_t \rho(t,u) 
\end{equation}
for $ (t,u) \in [0,T]\times(-1,1)$ with the boundary condition
$\psi(t,\pm 1) = 0$, $t\in [0,T]$.
\end{lemma}

\begin{demo}
Fix $t\in [0,T]$, for $h\neq 0$ such that $t+h\in [0,T]$ let us
introduce $\psi_h(t,u) := [\f(t+h,u)-\f(t,u)]/h$. Note that, by
(i)--(ii) in Theorem \ref{t:deq}, $\psi_h(t,\cdot)\in
C^{2}\left([-1,1]\right)$. By using \reff{Deqf}, we get
that $\psi_h$ solves  
\begin{equation}\label{dDeqfh}
\begin{array}{l}
{\displaystyle \vphantom{\Bigg(}
 \nabla\Big[
\frac{ \nabla \psi_h (t,u)}{ \nabla \f (t,u) \,  \nabla\f(t+h,u)} 
 \Big]
- \frac{e^{\f(t,u)}}
{\big( 1+ e^{\f(t,u)}\big) \, \big( 1+ e^{\f(t+h,u)}\big)} 
\: \frac{e^{h\,\psi_h(t,u)}-1}{h} 
}
\\
{\displaystyle 
\vphantom{\Bigg(}
\phantom{merda} 
= \; \frac{\rho(t+h,u)-\rho(t,u)}{h} }
\end{array}
\end{equation}
for $(t,u) \in [0,T]\times(-1,1)$ with the boundary condition 
$\psi_h(t,\pm 1) = 0$,  $t\in [0,T]$.

Multiplying the above equation by $\psi_h(t,u)$ and 
integrating in $du$, after using the inequality $x(e^x-1)\ge 0$ and 
an integration by parts, we get
$$
\begin{array}{l}
{\displaystyle 
\vphantom{\Bigg(}
\int_{-1}^1\!du \: 
\frac{ \left( \nabla \psi_h (t,u)\right)^2}
     { \nabla \f (t,u) \,  \nabla\f(t+h,u)}       
\le \left| \int_{-1}^1\!du \: \psi_h (t,u)\: 
        \frac{\rho(t+h,u)-\rho(t,u)}{h} \right|
}\\
{\displaystyle 
\vphantom{\Bigg(}
\phantom{merda} 
\le \e \int_{-1}^1\!du \: \psi_h (t,u) ^2 
+ \frac{1}{4\e} \int_{-1}^1\!du \: 
\Big( \frac{\rho(t+h,u)-\rho(t,u)}{h} \Big)^2
}
\end{array}
$$
where we used Schwartz inequality with $\e>0$.
Recalling the Poincar\'e inequality (with $f(\pm 1)=0$)
$$
\int_{-1}^{1} \!du \: f(u)^2 \le \frac{4}{\pi^2}   
\int_{-1}^{1} \!du \: f'(u)^2
$$
using \reff{sf'} and choosing $\e$ small enough we finally find 
$$
\limsup_{h\to 0}  \int_{-1}^1\!du \: \big[  \nabla \psi_h (t,u)\big] ^2 
\le C_2 \int_{-1}^1\!du \: \big[ \partial_t\rho (t,u) \big]^2
$$
for some constant $C_2$ depending only on $\rho_+$, $\rho_-$.

Hence, by Sobolev embedding, the sequence $\psi_h(t,\cdot)$ is
relatively compact in $C\big([-1,1]\big)$. By taking the limit $h\to
0$ in \reff{dDeqfh} it is now easy to show any limit point is a weak
solution of \reff{dDeqf}. By classical theory on the one--dimensional
elliptic problems, see e.g.\ \cite[IV, \S 2.1]{M}, there exists a
unique weak solution of \reff{dDeqf} which is in fact the classical
solution since $\partial_t \rho(t,\cdot) \in C\big([-1,1]\big)$. This
implies there exists a unique limit point $\psi(t,u)$ which is twice
differentiable w.r.t.\ $u$. The continuity of $t\mapsto \psi(t,\cdot)$
follows from the continuos dependence (in the $C^2\big([-1,1]\big)$
topology) of the solution of \reff{dDeqf} w.r.t.\ $\partial_t
\rho(t,\cdot)$ (in the $C\big([-1,1]\big)$ topology).
\end{demo}

The link between the boundary value problem \reff{Deq} and the
variational problem \reff{SsS} is established by the following theorem. 

\begin{theorem}
\label{t:s=s}
Let $S$ be the functional on $\mc{M}$ defined in \reff{SsS}. Then 
$S$ is bounded, convex and lower semi continuous on $\mc M$. 
Moreover, for each $\rho\in\mc M$, we have that
$S(\rho) = \mc{G} \big(\rho, F(\rho) \big)$ where $F(\rho)$ 
is the solution of \reff{DeqI}. 
\end{theorem}

\begin{demo}
For each $f\in \mc F$ we have that $\mc G (\cdot, f)$ is a convex
lower semi continuous functional on $\mc M$. 
Hence the functional $S(\cdot)$ defined in \reff{SsS}, being the
supremum of convex lower semi continuous functionals, is a convex lower
semi continuous functional on $\mc M$. Furthermore, by choosing
$f=\bar\rho$ in \reff{SsS} we obtain that $0\le S_0(\rho)\le
S(\rho)$. Finally, by using the concavity of $x\mapsto \log x$, 
Jensen's inequality, and  $f(\pm 1) = \rho_\pm$, 
we get that $\mc{G}(\rho,f)$ is bounded by some constant depending
only on $\rho_-$ and $\rho_+$. 

In order to show the supremum in \reff{SsS} is uniquely attained when
$f=F(\rho)$ solves \reff{DeqI}, it is convenient to make, as in Lemma
\ref{t:ddeq}, the change of variables $\f=\phi (f)$ defined by 
$\varphi(u)
:= \log \big\{f(u)/[1-f(u)]\big\}$. Note that
$f(u)=e^{\f(u)}/[1+e^{\f(u)}]$.  
We then need to show that the supremum of the functional
\begin{equation}\label{Grf}
\begin{array}{rcl}
{\displaystyle 
\!\!\!\! \vphantom{\Bigg(}
{\widetilde{ \mc{G}}}  (\rho,\f) 
}
&\!\!\! :=  \!\!\! &
{\displaystyle 
\mc{G}(\rho,\phi^{-1}(\f)) }
\\
\!\!\!\!
&\!\!\! =  \!\!\! &
{\displaystyle
\vphantom{\Bigg(}
\int_{-1}^1\!du \:
\bigg\{ \rho(u)\log\rho(u) + [1-\rho(u)]\log[1-\rho(u)]
}
\\
\!\!\!\!
&\!\!\! \!\!\! &
\vphantom{\Big(}
{\displaystyle \phantom{\int_{-1}^1\!du\:}
+[1-\rho(u)]\f(u) 
-\log\big[1+e^{\f(u)}\big]
+\log\frac{\f'(u)}{[\rho_+-\rho_-]/2} \bigg\}
}
\end{array}
\end{equation}
for $\f\in {\widetilde{\mc{F}}} := \phi(\mc{F}) 
= \big\{ \f \in C^1\big([-1,1]\Big) \, : \: 
\f(\pm 1)=\f_\pm\,,\: \f'(u) >0 \big\}$ is uniquely attained when 
$\f= \phi (F (\rho))$. We recall that $F(\rho)$ denotes the solution
of \reff{DeqI}.

Since the real functions $x\mapsto \log x$ and $x\mapsto
-\log\big(1+e^{x}\big)$ are strictly concave, for each $\rho\in\mc{M}$
the functional ${\widetilde{\mc{G}}}(\rho,\cdot)$ 
is strictly concave on ${\widetilde{\mc{F}}}$. 
Moreover it is easy to show  that ${\widetilde{\mc{G}}}(\rho,\cdot)$ 
is Gateaux
differentiable on ${\widetilde{\mc{F}}}$ with derivative given by
$$
\Big\langle 
\frac{\delta {\widetilde{\mc{G}}}(\rho,\f)}{\delta \f} , g \Big\rangle 
= \int_{-1}^1\!du \:
\left\{ \frac{g'(u)}{\f'(u)} + \bigg[ \frac{1}{1+e^{\f(u)}}
- \rho(u) \bigg] g(u)\right\}
$$
By standard convex analysis, see e.g.\ \cite[I, Prop. 5.4]{ET}, 
for any $\f\neq\psi\in{\widetilde{\mc{F}}}$ we have 
$$
{\widetilde{\mc{G}}}(\rho,\psi) <{\widetilde{\mc{G}}}(\rho,\f) +
\Big\langle \frac{\delta {\widetilde{\mc{G}}}(\rho,\f)}{\delta \f} , 
\psi-\f \Big\rangle 
$$
By noticing that $\delta {\widetilde{\mc{G}}} (\rho,\f) / 
{\delta \f} =0$ if
$\f$ solves \reff{Deqf} a.e.\ we conclude the proof that the 
supremum on ${\widetilde{\mc{F}}}$ of ${\widetilde{\mc{G}}}(\rho,\cdot)$ 
is uniquely attained when  $\f= \phi (F (\rho))$.
\end{demo}

\begin{remark}\label{rem46}
Given $\rho\in\mc M$, let us consider a sequence $\rho_n\in
C^2\big([-1,1]\big)\cap \mc M$ with $\rho_n(\pm 1)=\rho_\pm$, bounded
away from $0$ and $1$, which converges to $\rho$ a.e.  Then, by
dominated convergence and (ii) in Theorem \ref{t:deq}, we have
$S(\rho_n)=\mc{G}\big(\rho_n,F(\rho_n)\big) \longrightarrow
\mc{G}\big(\rho,F(\rho)\big)= S(\rho)$.
\end{remark}

\section{The quasi potential}
\label{s:5}
In this section we show that the quasi potential for the
one--dimensional boundary driven simple exclusion process, as defined
by the variational problem \reff{qp}, coincides with the functional 
$S(\rho)$ defined in \reff{SsS}. In the proof we shall
also construct an optimal path for the variational problem \reff{qp}.

Let us first recall the heuristic argument given in \cite{BDGJL}.
Taking into account the representation of the functional
$I_T(\pi|\bar\rho)$ given in Lemma \ref{s07}, to the variational
problem \reff{qp} is associated the Hamilton--Jacobi equation 
\begin{equation}
\label{HJ}
\frac 12 \bigg\langle \nabla \frac{\delta V}{\delta \rho}, 
\rho(1-\rho) \nabla \frac{\delta V}{\delta \rho} 
\bigg\rangle 
+ 
\bigg\langle \frac{\delta V}{\delta \rho}, \frac 12 \Delta \rho
\bigg\rangle = 0
\end{equation}
where $\nabla$ denotes the derivative w.r.t.\ the macroscopic
space coordinate $u\in [-1,1]$. We look for a solution in the form 
$$
\frac{\delta V}{\delta \rho} = 
\log \frac{\rho}{1-\rho} - \log \frac{f}{1-f}
$$
and obtain a solution of \reff{HJ} provided $f$ solves the boundary
value problem \reff{Deq}, namely $f=F(\rho)$.  
On the other hand, by Theorem \ref{t:s=s}, we have
$$
\begin{array}{rcl}
{\displaystyle 
\vphantom{\Bigg(}
\frac{\delta S(\rho)}{\delta \rho}}
& =&  
{\displaystyle \frac{\delta \mc{G}(\rho, f)}{\delta \rho} \bigg|_{f=F(\rho)}
+ \frac{\delta \mc{G}(\rho, f)}{\delta f} \bigg|_{f=F(\rho)}
\frac{\delta F(\rho)}{\delta \rho}
}
\\
&=& {\displaystyle
\vphantom{\Bigg(} 
\log \frac{\rho}{1-\rho} - \log \frac{F(\rho)}{1-F(\rho)} 
}
\end{array}
$$
since \reff{Deq} is the Euler--Lagrange equation for the variational
problem \reff{SsS}. We get therefore $V=S$ since we have
$V(\bar\rho)=S(\bar\rho)=0$. 

Let $\pi^*(t)=\pi^*(t,u)$ be the optimal path for the variational problem
\reff{qp} and define $\rho^*(t) := \pi^*(-t)$. By using a time reversal
argument, in \cite{BDGJL} it is also shown that $\rho^*(t)$ solves the
hydrodynamic equation associated to the adjoint process (whose
generator is the adjoint of $L_N$ in $L_2(d\mu^N)$) which takes the
form 
\begin{equation}\label{ahf}
\partial_t \rho^*(t) = 
-\frac 12 \Delta \rho^*(t) + \nabla\bigg( \rho^*(t) [1-\rho^*(t)] 
\nabla \frac {\delta S(\rho)}{\delta\rho}\Big|_{\rho=\rho^*(t)} 
\bigg)
\end{equation}

We will not develop here a mathematical theory of the Hamilton--Jacobi
equation \reff{HJ}.
We shall instead work directly with the variational problem \reff{qp},
making explicit computations for smooth paths and using approximation
arguments to prove that we have indeed $V=S$.  Of course, the
description of the optimal path will also play a crucial role.

To identify the quasi potential $V$ with the functional $S$ we shall
prove separately the lower bound $V\ge S$ and the upper bound $V\le
S$.  For this purpose we start with two lemmata, which connect $S$
defined in \reff{SsS} to the Hamilton--Jacobi equation \reff{HJ}, used
for both inequalities.  
The bound $V\ge S$ will then be proven by choosing the right test
field $H$ in \reff{j}.
To prove $V\le S$ we shall exhibit a path $\pi^*(t)=\pi^*(t,u)$ which
connects the stationary profile $\bar\rho$ to $\rho$ in some time
interval $[0,T]$ and such that $I_{T}(\pi^* | \bar \rho) \le S(\rho)$.
As outlined above, this path ought to be the time reversal of the
solution of the adjoint hydrodynamic equation \reff{ahf} with initial
condition $\rho$.  The adjoint hydrodynamic equation needs, however,
infinite time to relax to the stationary profile $\bar \rho$. We have
therefore to follow the time reversed adjoint hydrodynamic equation in
a time interval $[0,T_1]$ to arrive at some profile $\rho^*(T_1)$,
which is close to $\bar\rho$ if $T_1$ is large, and then interpolate,
in some interval $[T_1,T_1+T_2]$, between $\rho^*(T_1)$ and
$\bar\rho$.

\medskip
Recall that we are assuming $\rho_- < \rho_+$ and pick $\delta_0>0$ 
small enough for  $\delta_0 \le \rho_- <  \rho_+ \le 1- \delta_0$.
For $\delta\in(0,\delta_0]$ and $T>0$, we introduce
\begin{equation}
\label{dMd}
\mc{M}_\delta \;:=\; \{ \rho \in C^2\big([-1,1]\big)\, :\,
\rho(\pm 1) =\rho_\pm\,,\; \delta \le \rho(u) \le 1-\delta\,
\}
\;
\end{equation}
\begin{equation}
\label{dDd}
D_{T,\delta} \;:=\; \left\{ \pi \in C^{1,2} \big( [0,T]\times[-1,1]
\big) \,:\: \pi(t,\pm 1 ) =\rho_\pm \,,\; \delta \le \pi(t,u) \le
1-\delta \right\}
\end{equation}


\begin{lemma}
\label{dotS}
Let $\pi\in D_{T,\delta}$ and denote by $F(t,u) = F(\pi(t,\cdot))\,(u)$ the
solution of the boundary value problem (\ref{Deq}) 
with $\rho$ replaced by $\pi(t)$. Set 
\begin{equation}\label{dGt}
\Gamma (t,u)= \log \frac{\pi(t,u)}{1-\pi(t,u)} \; -\; 
\log \frac{F(t,u)}{1-F(t,u)} \; \cdot
\end{equation}
Then, for each $T\geq0$,
\begin{equation}\label{nome2}
S\big(\pi(T)\big) - S\big(\pi(0)\big) 
=  \int_0^T  \!dt\:  \langle \partial_t \pi(t), \Gamma(t)
\rangle \; .
\end{equation}
\end{lemma}

\begin{demo}
Note that $F(t,\cdot)$ is strictly increasing for any $t\in[0,T]$ and 
$F\in C^{1,2}\big([0,T]\times[-1,1]\big)$ by (iii) in
Theorem \ref{t:deq}. Moreover, since $F(t,\pm 1) = \rho_\pm$, we have
$\partial_t F(t,\pm 1)=0$.  
By Theorem \ref{t:s=s}, dominated convergence, 
an explicit computation, and an integration by parts, we get
\begin{equation*}
\begin{array}{l}
{\displaystyle 
\vphantom{\bigg(}
\frac {d}{dt} S\big(\pi(t)\big)
= \frac {d}{dt} \: \mc{G}\big(\pi(t), F(t) \big) 
}
\\
{\displaystyle 
\vphantom{\Bigg\{}
\phantom{\frac {d}{dt} S\big(\pi(t)\big)}
= \Big\langle \partial_t \pi(t), \Gamma(t) \Big\rangle
+ 
\Big\langle \partial_t F(t), \frac{1-\pi(t)}{1-F(t)} -\frac{\pi(t)}{F(t)}
\Big\rangle
 + \Big\langle \frac{1}{\nabla F(t)}, \partial_t \nabla F(t)\Big \rangle 
}
\\
{\displaystyle 
\phantom{\frac {d}{dt} S\big(\pi(t)\big)} =
\Big\langle \partial_t \pi(t), \Gamma(t) \Big\rangle
+ \Big\langle \partial_t F(t), \frac{F(t)-\pi(t)}{F(t)[1-F(t)]} + 
\frac{\Delta F(t)}{\big(\nabla F(t)\big)^2} \Big\rangle 
}
\end{array}
\end{equation*}
The lemma follows by noticing that the last term above vanishes by 
\reff{Deq}.
\end{demo}

\begin{lemma}
\label{ham-jac}
Let $\rho\in \mc{M}_\delta$,
denote
by $F(u) = F(\rho)\,(u)$ the solution of the boundary value problem
\reff{Deq}, and set
$$
\Gamma (u) =  \log \frac{\rho(u)}{1-\rho(u)} -
\log \frac{F(u)}{1-F(u)}\; \cdot
$$
Then,
\begin{equation}
\label{h.1}
\big\langle \rho (1-\rho), \big(\nabla \Gamma\big)^2  \big\rangle
\; +\; \big\langle \Delta \rho , \Gamma  \big\rangle \; =\;  0 \; .
\end{equation}
\end{lemma}

\begin{demo}
Note that $F\in \mc{M}_{\delta}$ by Theorem \ref{t:deq}.
After an integration by parts and simple algebraic manipulations
\reff{h.1} is equivalent to
\begin{equation}
\label{h.2}
- \Big\langle \nabla \rho, \frac{\nabla F}{F(1-F)} \Big\rangle
\; +\;  \Big\langle \rho (1-\rho) , \Big( 
\frac{\nabla F}{F(1-F)}
\Big)^2 \Big\rangle \;=\;0 \;.
\end{equation}
We rewrite the first term on the left hand side as
$$
- \Big\langle \nabla F, \frac{\nabla F}{F(1-F)} \Big\rangle 
- \Big\langle \nabla (\rho-F) , \frac{\nabla F}{F(1-F)}\Big\rangle 
$$
which, by an integration by parts, is equal to
$$
- \Big\langle \nabla F, \frac{\nabla F}{F(1-F)}\Big\rangle 
+ \Big\langle\rho-F,  \frac{\Delta F}{F(1-F)}  - \frac{(1-2F) 
\big(\nabla F\big)^2}{[F(1-F)]^2} \Big\rangle \; .
$$
Hence, the left hand side of \reff{h.2} is given by
\begin{equation*}
\begin{array}{l}
{\displaystyle 
\Big\langle \rho-F,  \frac{\Delta F}{F(1-F)} \Big\rangle 
- \Big\langle \frac{\big(\nabla F\big)^2}{[F(1-F)]^2} , 
(\rho - F)^2 \Big\rangle  
}
\\
{\displaystyle 
\qquad = \; \Big\langle \frac{\rho-F}{F(1-F)} \; , \;
\Delta F-(\rho-F) \frac{\big(\nabla F\big)^2}{F(1-F)} \Big\rangle  \; =0
}
\end{array}
\end{equation*}
thanks to
\reff{Deq}.
\end{demo}

Note that, for smooth paths, Lemma \ref{dotS} identifies, 
in the sense given by equation \reff{nome2}, $\Gamma$ as the
derivative of $S$. Lemma \ref{ham-jac} then states that this derivative
satisfies the Hamilton--Jacobi equation \reff{HJ}.

\subsection{Lower bound}
We can now prove the first relation between the quasi potential $V$
and the functional $S$.

\begin{lemma}
\label{qp>s} 
For each $\rho\in \mc M$ we have $V(\rho)\ge S(\rho)$.
\end{lemma}

\begin{demo}
In view of the variational definition $V$, to prove the lemma we need
to show that $S(\rho) \le I_T(\pi |\bar\rho)$ for any $T>0$ and any
path $\pi\in D\big([0,T];\mc{M}\big)$ which connects the 
stationary profile $\bar\rho$ to $\rho$ in the time interval $[0,T]$: 
$\pi(0)=\bar\rho$, $\pi(T)=\rho$.
  
Fix such a path $\pi$ and let us assume first that $\pi\in D_{T,\delta}$. 
Denote by $F(t) = F(\pi(t))$ the solution of the elliptic problem
(\ref{Deq}) with $\pi(t)$ in place of $\rho$.  In view of the
variational definition of $I_T(\pi |\bar\rho)$ given in \reff{eq:3},
to prove that $S(\rho) \le I_T(\pi |\bar\rho)$ it is enough to exhibit
some function $H \in C^{1,2}_0 ([0,T] \times [-1,1])$ for which
$S(\rho) \le J_{T,H,\bar{\rho}} (\pi)$.  We claim that $\Gamma$
given in \reff{dGt} fulfills these conditions.

We have that $\Gamma\in C^{1,2}_0 ([0,T] \times [-1,1])$ because: 
$\pi \in D_{T,\delta}$ by hypothesis, 
$F\in C^{1,2} ([0,T] \times [-1,1])$ by (iii) in 
Theorem \ref{t:deq}, $\Gamma(t,\pm 1) =0 $ since 
$\pi(t,\cdot)$ and $F(t,\cdot)$ satisfy the same boundary 
conditions. 
Recalling (\ref{j}) we get, after integration by parts,
\begin{equation*}
\begin{array}{rcl}
{\displaystyle 
\vphantom{\Bigg(}
J_{T,\Gamma,\bar{\rho}} (\pi)} &= & 
{\displaystyle 
\int_0^T \!dt\: 
\big\langle 
\partial_t \pi (t), \Gamma(t) \big\rangle 
}
\\
&&
{\displaystyle 
\vphantom{\Bigg(}
- \; \frac 12  \int_0^T \!dt\: \left[\,
\big\langle  \Gamma(t) , \Delta \pi(t)  \big\rangle 
+\big\langle  \pi(t)[1-\pi(t)], [\nabla \Gamma(t)]^2
\big\rangle  \, \right]\; .
}
\end{array}
\end{equation*}
By Lemmata \ref{dotS} and \ref{ham-jac} we then have 
$J_{T,\Gamma,\bar{\rho}} (\pi) = S(\rho)$.

Up to this point we have shown that $S(\rho) \le I_T(\pi |\bar\rho)$
for smooth paths $\pi$ bounded away from $0$ and $1$. In order to
obtain this result for general paths, we just have to recall the
approximations performed in the proof of the lower bound of the large
deviation principle. Fix a path $\pi$ with finite rate function:
$I_T(\pi|\bar\rho) < \infty$. 
In Section \ref{s:lb} we proved that  there exists a sequence
$\{\pi_n ,\, n\ge 1\}$ of smooth paths such that $\pi_n$ converges to
$\pi$ and $I_T(\pi_n|\bar\rho)$ converges to $I_T(\pi|\bar\rho)$. Let
$\tilde \pi_n$ be defined by $(1-n^{-1})\pi_n + n^{-1}\bar\rho$. Since
$\pi_n$ converges to $\pi$, $\tilde \pi_n$ converges to $\pi$. By
lower semi continuity of the rate function, $I_T(\pi|\bar\rho) \le
\liminf_{n\to\infty} I_T(\tilde \pi_n|\bar\rho)$. On the other hand,
by convexity, $I_T(\tilde \pi_n|\bar\rho) \le (1-n^{-1}) 
I_T(\pi_n|\bar\rho) + n^{-1} I_T(\bar\rho|\bar\rho) = (1-n^{-1})
I_T(\pi_n|\bar\rho)$ so that $\limsup_{n\to\infty} I_T(\tilde
\pi_n|\bar\rho) \le I_T(\pi |\bar\rho)$. Since $\tilde \pi_n$ belongs
to $D_{T,\delta}$ for some $\delta = \delta_n>0$, each path $\pi$ with
finite rate function can be approximated by a sequence $\tilde\pi_n$ in
$D_{T,\delta_n}$, for some set of strictly positive parameters
$\delta_n$, and such that $I_T(\pi|\bar\rho)=\lim_{n}
I_T(\tilde\pi_n|\bar\rho)$. 
Therefore, by the result on smooth paths and the
lower semi continuity of $S$, we get
$$
I_T(\pi|\bar\rho) \; =\; \lim_{n} I_T(\tilde\pi_n|\bar\rho) \ge 
\liminf_{n} S\big(\tilde\pi_n(T) \big) \ge S(\pi(T)) 
$$ 
which concludes the proof of the lemma. 
\end{demo}

\subsection{Upper bound}
The following lemma explains which is the right candidate for the 
optimal path for the variational problem \reff{qp}.

\begin{lemma}
\label{I=I*+S}
Fix $\delta\in (0,\delta_0]$, a profile $\alpha \in \mc M_\delta$,  
and a path $\pi \in D_{T,\delta}$ with finite rate function, 
$I_T(\pi|\alpha)<\infty$.
Denote by $F(t,u) = F(\pi(t,\cdot))\,(u)$ the solution
of the boundary value problem (\ref{Deq}) with $\rho$ replaced by
$\pi(t)$. Then there exists a function $K\in\mathcal{H}_1(\pi)$ such
that $\pi$ is the weak solution of
\begin{equation}
\label{f07bis}
\!\!\!
\left\{
\begin{array}{ll}
{\displaystyle
\partial_t \pi = - \frac 12  \Delta \pi 
+  \nabla \Big( 
\pi (1-\pi)  \nabla  \big[ \log \frac{ F}{1-F} + K \big] 
\Big)     
}
&\!\! (t,u)\in [0,T]\times (-1,1)
\\
{\displaystyle
\vphantom{\Big\{}
\pi(t, \pm 1) = \rho_\pm 
}
&\!\!
t\in [0,T]
\\
{\displaystyle
\vphantom{\Big\{}
\pi(0, u) = \alpha (u) 
}
&\!\!
u\in [-1,1]
\end{array}
\right.
\end{equation}
Moreover, 
\begin{equation}\label{I=}
I_T(\pi|\alpha)=  S(\pi(T)) -  S(\alpha)  + 
\frac 12 \int_0^T \!dt\, 
\big\langle\pi(t) [1-\pi(t)] , [ \nabla K(t)]^2 \big\rangle
\end{equation}
\end{lemma}

The optimal path for the variational problem \reff{qp} will be
obtained by taking a path $\pi^*$ for which the last term on the
right hand side of the identity \reff{I=} (which is positive)
vanishes, namely for a path $\pi^*$ which satisfies \reff{f07bis} with
$K=0$. Then $\rho^*(t)=\pi^*(-t)$ will be a solution of \reff{ahf}.

\begin{demo} 
Denote by $H$ the function in $\mathcal{H}_1(\pi)$ introduced in
Lemma \ref{s07},
let $\Gamma$ as defined in (\ref{dGt}), and set
$K:=\Gamma-H$.
Note that $K$ belongs to $\mathcal{H}_1(\pi)$ because: 
$\pi\in D_{T,\delta}$ by hypothesis, 
$F\in C^{1,2}\big([0,T]\times [-1,1]\big)$ by Theorem \ref{t:deq}, and 
$\Gamma(t,\pm 1)= 0$.
Then \reff{f07bis} follows easily from \reff{f07}.
To prove the identity \reff{I=},
replace in (\ref{nome2}) $\partial_t \pi(t)$ by the right 
hand side of the differential equation
in \reff{f07bis}. After an integration by parts
we obtain
\begin{equation*}
\begin{array}{rcl}
{\displaystyle 
S(\pi(T)) - S(\alpha) 
}
&=&  
{\displaystyle  
\int_0^T \!dt 
\: \bigg\{
\frac 12  \big\langle \Gamma (t),  \Delta \pi(t) \big\rangle 
+\big\langle \pi(t)[ 1-\pi(t)], [ \nabla \Gamma(t)]^2\big\rangle
}
\\
&&
{\displaystyle  
\phantom{\int_0^T \!dt \bigg\{}
-\;
\big\langle \pi(t)[1-\pi(t)],  \nabla K(t) \,  \nabla
\Gamma(t) \big\rangle
\bigg\}
}
\\
&=&
{\displaystyle
\int_0^T \!dt \Big\langle
\pi(t)[1-\pi(t)], \frac{1}{2}[\nabla\Gamma(t)]^2-\nabla\Gamma(t)\nabla K(t)
\Big\rangle}
\end{array}
\end{equation*}
where we used Lemma \ref{ham-jac}.
Recalling $K=\Gamma-H$, we thus obtain
\begin{equation*}
\begin{array}{l}
{\displaystyle
S(\pi(T)) - S(\alpha) + 
\frac 12 \int_0^T \!dt \, 
\big\langle \pi(t)[1-\pi(t)] , [ \nabla K(t)]^2\big\rangle  
}
\\
\qquad\quad
{\displaystyle
=\;  \frac 12 \int_0^T \!dt \, 
\big\langle \pi(t) [1-\pi(t)], [ \nabla H(t) ]^2\big\rangle 
} 
\end{array}
\end{equation*}
which concludes the proof of the lemma in view of \reff{f03}.
\end{demo}

We write more explicitly the adjoint hydrodynamic equation
\reff{ahf}. 
In the present paper, we shall use it only to describe a particular
path which will be shown to be the optimal one.
For $\rho\in\mc{M}$, consider the non local differential equation
\begin{equation}
\label{adjhyd}
\left\{
\begin{array}{ll}
{\displaystyle
\partial_t \rho^* = \frac 12  \Delta \rho^* 
- \nabla \Big( \rho^* (1-\rho^*) 
\nabla \log \frac{ F}{1-F} \Big)    
}
&
(t,u)\in(0,\infty)\times[-1,1]
\\
{\displaystyle 
\vphantom{\Big\{}
F(t,u) = F(\rho^*(t,\cdot))\,(u) 
}
& (t,u)\in(0,\infty)\times[-1,1]
\\
{\displaystyle
\vphantom{\Big\{}
\rho^*(t, \pm 1) = \rho_\pm 
}
&
t\in(0,\infty)
\\
{\displaystyle
\vphantom{\Big\{}
\rho^*(0,u) = \rho (u)
}
& u\in [-1,1]
\end{array}
\right.
\end{equation}
where we recall that $F(t,u) = F (\rho^*(t,\cdot))\,(u)$ means that
$F(t,u)$ has to be obtained from $\rho^*(t,u)$ by solving (\ref{DeqI}) 
with $\rho(u)$ replaced by $\rho^*(t,u)$.  
Since $\nabla \log [F/(1-F)] >0$, in (\ref{adjhyd}) there is a
positive drift to the right. Let us describe 
how it is possible to construct the solution of \reff{adjhyd}.

\begin{lemma}
\label{st1}
For $\rho\in\mc{M}$ let $\Phi(t)$ be the solution of the heat equation
\reff{hF} and define $\rho^*=\rho^*(t,u)$ by \reff{eq:4}. Then 
$\rho^*\in C^{1,2}\big((0,\infty)\times[-1,1]\big)\cap 
C\big([0,\infty);\mc{M}\big)$ and solves \reff{adjhyd}. 
Moreover, if $\delta \le \rho(u) \le 1-\delta$ a.e.\ for some
$\delta>0$, there exists $\delta'=\delta'(\rho_-,\rho_+,\delta)\in(0,1)$, 
for which $\delta' \le \rho^*(t,u) \le 1-\delta'$ for any 
$(t,u) \in (0,\infty) \times [-1,1]$. 
\end{lemma}

\begin{demo}
Let $F(u)=F(\rho)\,(u)$, then, by Theorem \ref{t:deq}, $F\in
C^1\big([-1,1]\big)$ and, by Lemma \ref{t:sch}, there is a constant
$C\in(0,\infty)$ depending only on $\rho_-$, $\rho_+$
such that $C^{-1} \le F'(u) \le C$ for any $u\in [-1,1]$.
Since $\Phi(t,u)$ solves \reff{hF}, there exists
$C_1=C_1(\rho_-,\rho_+)\in (0,\infty)$ such that $C_1^{-1} \le
( \nabla \Phi)(t, u) \le C_1$ for any 
$(t,u)\in [0,\infty)\times [-1,1]$. 
Moreover,
$\Phi(t,\pm 1) = \rho_{\pm}$ so that $\Delta \Phi(t,\pm 1) = 2
\partial_t \Phi(t,\pm 1) =0$.  Hence, $\rho^*$ defined by (\ref{eq:4})
satisfies the boundary condition $\rho^* (t, \pm 1) = \Phi^*(t,\pm 1)
= \rho_{\pm}$. 
Furthermore $\rho^*\in C^{1,2}\big((0,\infty)\times [-1,1]\big)$ .

For the reader's convenience, we reproduce below from \cite[Appendix
B]{BDGJL} the proof that $\rho^*(t,u)$, as defined in \reff{eq:4},
solves the differential equation in (\ref{adjhyd}). From \reff{eq:4}
we get that 
\begin{equation*}
\frac{ \rho^*(1-\rho^*)}{ \Phi(1-\Phi)} = 
1 + (1-2\Phi) \frac{\Delta \Phi}{ \big(\nabla \Phi\big)^2} 
- \Phi(1-\Phi) \frac{\big(\Delta \Phi\big)^2}{\big(\nabla \Phi\big)^4}
\end{equation*}
recalling \reff{hF}, by a somehow tedious computation of the partial
derivatives which we omit, we get
\begin{equation*}
\left( \partial_t - \frac 12 \Delta \right) 
\left[ \Phi (1-\Phi) \frac {\Delta \Phi}{\big(\nabla \Phi\big)^2} \right]
= - \nabla \Big( \frac{ \rho^*(1-\rho^*)}{ \Phi(1-\Phi)} \nabla \Phi \Big)
\end{equation*}
from which, by using again \reff{eq:4}, we see that $\rho^*$
satisfies the differential equation in \reff{adjhyd}.

To conclude the proof of the lemma, notice that $\rho^*$ is the
solution of  
\begin{equation*}
\left\{
\begin{array}{l}
{\displaystyle 
\partial_t \rho^* = \frac 12 \Delta \rho^* 
-  \nabla \{ \rho^* (1-\rho^*)  \nabla H \}
}\; , \\
{\displaystyle  
\vphantom{\Big\{}
\rho^*(t, \pm 1) = \rho_\pm
} \; , \\
{\displaystyle 
\vphantom{\Big\{}
\rho^*(0, \cdot) = \rho (\cdot)
}
\; ,
\end{array}
\right.
\end{equation*}

for some function $H$ in $C^{1,1}\big([0,\infty)\times [-1,1] \big)$
for which $ \nabla H$ is uniformly bounded. Though $H$ does not vanish
at the boundary, we may use a weakly asymmetric boundary driven
exclusion process to prove the existence of a weak solution $\lambda
(t,u)$, in the sense of Subsection \ref{s:3.4}, which takes values in
the interval $[0,1]$. Since $ \nabla H$ is bounded, the usual $H_{-1}$
method gives uniqueness so that $\lambda = \rho^*$ and $0\le \rho^*
\le 1$.  In particular $\rho^*\in C\big([0,\infty); \mc{M} \big)$.

Assume now that $\delta \le \rho \le 1-\delta$ for some $\delta >0$.
Fix $t>0$ and assume that $\rho^*(t, \cdot)$ has a local maximum at
$-1<u_0<1$. Since $\rho^*$ is a smooth solution of (\ref{adjhyd}), a
simple computation gives that at $(t,u_0)$
$$
(\partial_t \rho^*) \;=\; \frac 12  \Delta \rho^* 
- \frac{\rho^*  (1-\rho^*) (\nabla F)^2}{F^2(1-F)^2}  
(\rho^* + F -1) 
$$
because $( \nabla \rho^*)(t,u_0)=0$ and $\Delta \log \{F/1-F\} =
(\nabla F)^2 (\rho^* + F -1)/F^2(1-F)^2$. Since $u_0$ is a local maximum,
$\Delta \rho^* \le 0$. On the other hand, assume that $\rho^*(t,u_0) >
1-\rho_-$, in this case, since $\rho_- \le F$, $\rho^* + F -1 >0$ so
that $\partial_t \rho^* <0$.  In the same way we can conclude that 
$(\partial_t \rho^*)(t,u_1) >0$ if $u_1$ is a minimum of $\rho^*(t,\cdot)$
and $ \rho^* (t,u_1) \le 1-\rho_+$. These two estimates show that
$\min\{\delta , 1-\rho_+, \rho_-\} \le  \rho^*(t,u) \le \max \{1-\delta,
1-\rho_-, \rho_+ \}$, which concludes the proof of the lemma. 
\end{demo}

We now prove that the solution of (\ref{adjhyd}), as constructed in
Lemma \ref{st1}, converges, as $t\to\infty$, to $\bar\rho$
uniformly with respect to the initial datum $\rho$. We use below the
usual notation $\|f\|_\infty := \sup_{u\in [-1,1]} |f(u)|$.

\begin{lemma}
\label{convergence!}
Given $\rho\in\mc{M}$, let $\rho^*(t)=\rho^*(t,u)$ be the solution
\reff{adjhyd}. Then,
$$
\lim_{t\rightarrow \infty}  \sup_{\rho \in\mc{M}} \; 
\big\Vert \rho^*(t) -\bar\rho \big\Vert_\infty \; =\; 0 \; .
$$
\end{lemma}

\begin{demo}
Let us represent the solution $\Phi(t)$ of \reff{hF} in the form
$\Phi(t,u)=\bar\rho(u) + \Psi (t,u)$. Then $\Psi(t) = P^0_t \Psi(0)$
where $ P^0_t$ is the semigroup generated by $(1/2) \Delta^0$, with
$\Delta^0$ the Dirichlet Laplacian on $[-1,1]$. Since $\Psi(0)= F(\rho
) -\bar\rho$ and since the solution $F(\rho)$ of (\ref{DeqI}) as well
as $\bar \rho$ are contained in the interval $[\rho_-, \rho_+]$, we
have that $\|\Psi(0)\|_\infty \le |\rho_+ - \rho_-|<1$.  
Therefore, by standard heat kernel estimates,
$$
\lim_{t\to\infty} \sup_{\rho \in\mc{M}} \; \Big\{ 
\| \Psi(t) \|_\infty +\|\nabla \Psi(t) \|_\infty  
+ \|\Delta \Psi(t) \|_\infty \Big\} \;=\; 0
$$
the lemma follows recalling that, by Lemma \ref{st1},  $\rho^*(t)$
is given by \reff{eq:4}. 
\end{demo}

Lemma \ref{convergence!} shows that we may join a profile $\rho$ in
$\mc M$ to a neighborhood of the stationary profile by using the
equation \reff{adjhyd} for a time interval $[0,T_1]$ which at the same
time regularizes the profile.
On the other hand, from Lemma \ref{I=I*+S} we shall deduce
that this path pays $S(\rho)-S(\rho^*(T_1))$.  It thus remains to
connect $\rho^*(T_1)$, which is a smooth profile close to the
stationary profile $\bar\rho$ for large $T_1$, to $\bar\rho$. In the
next lemma we show this can be done by paying only a small price.  We
denote by $\|\cdot\|_2$ the norm in $L_2\big([-1,1],du\big)$.

\begin{lemma}
\label{inter}
Let $\alpha \in \mc{M}_{\delta_0}$ be a smooth profile such that
$\|\alpha - \bar\rho\|_\infty \le \delta_0/(16)$. Then there exists a
smooth path $\hat\pi(t)$, $t\in [0,1]$ with $\delta_0/2 \le\hat\pi\le
1- \delta_0/2$, namely $\hat\pi\in D_{1,\delta_0/2}$, with
$\hat\pi(0)=\bar\rho$, $\hat\pi(1)=\alpha$ and a constant
$C=C(\delta_0)\in (0,\infty)$ such that 
$$
I_1(\hat\pi|\bar\rho) \le C \| \alpha -\bar\rho\|_2^2
$$
In particular $V(\alpha) \le C \| \alpha -\bar\rho\|_2^2$.
\end{lemma}

We remark that by using the ``straight path'' $\hat\pi (t) =
\bar\rho\, (1-t) + \alpha \,t$ one would get a bound in terms of the
$H_{1}$ norm of $\alpha-\bar\rho$. Below, by choosing a more clever
path, we get instead a bound only in term of the $L_2$ norm.

\smallskip
\begin{demo}
Let $(e_k,\lambda_k)$, $k\ge 1$ be the spectral basis for
$-(1/2)\Delta^0$, where $\Delta^0$ is the Dirichlet Laplacian on
$[-1,1]$, namely $\{e_k\}_{k\ge 1}$ is an complete orthonormal system
in $L_2([-1,1],du)$ and $ -(1/2)\Delta^0 e_k =\lambda_k e_k$.  
Explicitly we have $e_k(u) = \cos ( k\pi u /2 )$ and
$\lambda_k = k^2 \pi^2 / 8$. We claim that the path
$\hat\pi(t)=\hat\pi(t,u)$, $(t,u)\in [0,1]\times [-1,1]$ given by
\begin{equation}
\label{pih}
\hat\pi(t) =  \bar\rho + \sum_{k=1}^\infty 
\frac{e^{\lambda_k t} -1}{e^{\lambda_k} -1} 
\langle \alpha -\bar\rho, e_k\rangle e_k
\end{equation}
fulfills the conditions stated in the lemma. 

It is immediate to check that $\hat\pi(0)=\bar\rho$, $\hat\pi(1)=\alpha$
and $\hat\pi(t,\pm 1)=\rho_\pm$. Furthermore, by the smoothness assumption on
$\alpha$, we get that $\hat\pi\in C^{1,2}\big([0,1]\times
[-1,1]\big)$. In order to show that $\delta_0/2 \le\hat\pi\le 1- \delta_0/2$,
let us write $\hat\pi(t)=\bar\rho + q(-t)$, then $q(t)=q(t,u)$,
$(t,u)\in [-1,0]\times[-1,1]$ solves
\begin{equation*}
\left\{
\begin{array}{l}
{\displaystyle 
\partial_t q(t) = \frac 12 \Delta q(t) + g 
}
\\
{\displaystyle 
\vphantom{\Big(}
q(t,\pm 1) = 0 
}
\\
{\displaystyle 
q(-1,u) = \alpha(u) -\bar\rho(u)
}
\end{array}
\right.
\end{equation*}
where $g=g(u)$ is given by
$$
g = - \sum_{k=1}^\infty \frac{\lambda_k}{e^{\lambda_k} -1} 
\langle \alpha -\bar\rho, e_k\rangle e_k
$$

Let us denote by $\| g\|_{H_1}:= \| g' \|_2$ the $H_1$ norm
in $[-1,1]$; a straightforward computation shows
\begin{equation*}
\begin{array}{rcl}
{\displaystyle 
\| g\|_{H_1}^2 
}
&=&  
{\displaystyle \vphantom{\Big\{}
\sum_{k=1}^\infty  2 \lambda_k
\Big( \frac{\lambda_k}{e^{\lambda_k} -1} \Big)^2 
\langle \alpha -\bar\rho, e_k\rangle^2
\le
\frac{8}{\lambda_1}  \sum_{k=1}^\infty  
 \langle \alpha -\bar\rho, e_k\rangle^2
}
\\
&\le& 
{\displaystyle 
\frac{8}{\lambda_1} \| \alpha -\bar\rho \|_2^2 
\le \Big( \frac{8}{\pi} \Big)^2 \, 2 \Big( \frac{\delta_0}{16} \Big)^2 
=  \frac{1}{2\pi^2} \delta_0^2
}
\end{array}
\end{equation*}
where we used that, for $\lambda>0$, we have $e^\lambda -1 \ge
\lambda^2/2$.

Let $P_t^0 = \exp\{t \Delta^0/2\}$ be the heat semigroup on $[-1,1]$;
since $\|g\|_\infty \le \sqrt{2} \|g\|_{H_1}$ we have
$$
\sup_{t\in[-1,0]} \|q(t)\|_\infty  = 
\sup_{t\in[-1,0]} \Big\| P_{t+1}^0 (\alpha -\bar\rho) 
+ \int_{-1}^t \!ds \: P^0_{t-s} g \Big\|_\infty  \le
\frac{\delta_0}{16} +  \frac{1}{\pi} \delta_0  \le \frac{7}{16} \delta_0
$$
so that $\hat\pi \in D_{1,\delta_0/2}$.

We shall estimate $I_1(\hat\pi|\bar\rho)$ by using the
representation given in Lemma \ref{s07}. To this end, let us define
$h=h(t,u) \in C\big([0,1]\times[-1,1]\big)$ by $h := - \partial_t
\hat\pi + (1/2)\Delta \hat\pi$ and let $H=H(t,u)$ be the solution of
\begin{equation*}
\left\{
\begin{array}{l}
{\displaystyle 
 \nabla\big( \hat\pi [1-\hat\pi]  \nabla H \big) = h 
}
\\
{\displaystyle 
\vphantom{\bigg(}
H(t,\pm 1) = 0
}
\end{array}
\right.
\end{equation*}
so that $\hat\pi$ solves \reff{f07} with $H$ as above which
belongs to $\mc{H}_1(\hat\pi)$. 

Let us denote by $\|\cdot\|_{H_{-1}}$ the usual negative Sobolev norm
in $[-1,1]$, namely
$$
\| h \|_{H_{-1}}^2 := \sup_{f\neq 0, f(\pm 1)=0} \frac 
{\langle f, h \rangle^2}{\langle  \nabla f,  \nabla f \rangle}
=\sum_{k=1}^\infty \frac{1}{2\lambda_k} \langle h,e_k\rangle^2
$$ 
By using that $\hat\pi [1-\hat\pi] \ge (\delta_0/2)^2$ 
a simple computations shows
$$
\int_0^1\!dt \: \big\langle  
\hat\pi(t) [1-\hat\pi(t)] , \big( \nabla H(t)\big)^2 \big\rangle
\le  \frac {4}{\delta_0^2} \int_0^1\!dt \: 
\| h(t) \|_{H_{-1}}^2
$$
By using the explicit expression for $\hat\pi$ we get
$$
h(t) = - \sum_{k=1}^\infty \lambda_k 
\frac{2 e^{\lambda_k t} -1}{e^{\lambda_k}-1} 
\langle \alpha -\bar\rho, e_k\rangle e_k 
$$
hence, by a direct computation,
$$
\begin{array}{rcl}
{\displaystyle 
\|h(t) \|_{H_{-1}}^2
}
&=& 
{\displaystyle 
\sum_{k=1}^\infty \frac{1}{2\lambda_k} 
\Big( \lambda_k \frac{2 e^{\lambda_k t} -1}{e^{\lambda_k}-1} \Big)^2 
\langle \alpha -\bar\rho, e_k\rangle^2
}
\\
&\le&
{\displaystyle \vphantom{\bigg(}
\sum_{k=1}^\infty  8 \lambda_k e^{2\lambda_k (t-1)}
\langle \alpha -\bar\rho, e_k\rangle^2
}
\end{array}
$$
where we used that for $\lambda\ge \lambda_1$ we have $e^\lambda\ge 2 $. 
We thus get
$$
I_1(\hat\pi|\bar\rho) \le 
\frac {2}{\delta_0^2} \int_0^1\!dt \: \|h(t) \|_{H_{-1}}^2
\le \frac {8}{\delta_0^2} 
\sum_{k=1}^\infty  \langle \alpha -\bar\rho, e_k\rangle^2
=\frac {8}{\delta_0^2} \|\alpha-\bar\rho\|_2^2
$$
which concludes the proof of the lemma.
\end{demo}

We can now prove the upper bound for the quasi potential and conclude
the proof of Theorem \ref{s02}.

\begin{lemma}
\label{qp<s}
For each $\rho \in \mc M$, we have  $V(\rho) \le S(\rho)$.
\end{lemma}

\begin{demo}
Fix $0<\varepsilon < \delta_0/(32)$, 
$\rho\in \mc M$ and let $\rho^*(t,u)$ be the solution of
\reff{adjhyd} with initial condition $\rho$. By Lemma
\ref{convergence!} there exists $T_1=T_1(\varepsilon)$ such that
$\| \rho^*(t) -\bar\rho \|_\infty < \varepsilon$ for any $t\ge T_1$.
Let $\alpha := \rho^*(T_1)$ and let $\hat\pi$ be the path which
connects $\bar\rho$ to $\alpha$ in the interval $[0,1]$ constructed in
Lemma \ref{inter}.

Let $T:=T_1+1$ and $\pi^*(t)$, $t\in [0,T]$ the path
\begin{equation}\label{pi*}
\pi^* (t) \; =\; 
\left\{ 
  \begin{array}{ll}
\hat\pi (t) & \text{for $0\le t\le 1$} \\
\rho^* (T - t)  & \text{for $1\le t\le T$}
  \end{array}
\right.
\end{equation}

By Remark \ref{rem46}, given $\rho\in\mc{M}$ as above, we can find 
a sequence $\{\rho_n ,\, n\ge 1\}$ with $\rho_n\in \mc M_{\delta_n}$
for some $\delta_n>0$ converging to $\rho$ in $\mc M$ and  
such that $S(\rho_n)$ converges to $S(\rho)$.
Let us denote by $\rho^{n,*}$ the solution of \reff{adjhyd} with initial
condition $\rho_n$ and set  
\begin{equation}\label{pi*n}
\pi^{n,*} (t) \; =\; 
\left\{ 
  \begin{array}{ll}
\hat{\pi}^{n,*} (t) & \text{for $0\le t\le 1$} \\
\rho^{n,*} (T - t)  & \text{for $1\le t\le T$}
  \end{array}
\right.
\end{equation}
where $\hat{\pi}^{n,*} (t)$ is the path joining $\bar\rho$ to
$\alpha_n :=\rho^{n,*}(T_1)$ in the time interval $[0,1]$ constructed in
Lemma \ref{inter}. We claim that the path $\pi^{n,*}$ defined above
converges in $D\big([0,T], \mc M\big)$ to $\pi^*$, as defined in
\reff{pi*}. Before proving this claim, we conclude the proof of
the lemma.  

By the lower semi continuity of the functional $I_T(\cdot|\bar\rho)$ on 
$D\big([0,T], \mc M\big)$ we have
\begin{equation}\label{4}
I_T (\pi^* | \bar \rho) 
\le \liminf_{n} I_T (\pi^{n,*} | \bar \rho)
\end{equation}
On the other hand, by definition of the rate function and its
invariance with respect to time shifts we get
\begin{equation}
\label{eq:5}
I_T \big( \pi^{n,*}\big|\bar \rho \big) 
=
I_1 \big(\hat{\pi}^{n,*} \big| \bar \rho \big) + 
I_{T_1} \big( \rho^{n,*}(T_1-\cdot) \big| \rho^{n,*}(T_1)\big)
\end{equation}

By Theorem \ref{t:deq}, $F_n:=F(\rho_n)$ converges to $F=F(\rho)$ in
$C^1\big([-1,1]\big)$ so that $\Phi_n(t)$, the solution of \reff{dF}
with initial condition $F_n$, converges to $\Phi(t)$ in 
$C^2\big([-1,1]\big)$ for any $t>0$. Hence, by \reff{eq:4},
$\rho^{n,*}(T_1)$ converges to $\rho^*(T_1)$ in $C\big([-1,1]\big)$.
Recalling that 
$\| \rho^*(T_1) - \bar\rho\|_\infty < \varepsilon \le \delta_0/(32)$, 
we can find $N_0=N_0(\delta_0)$ such that for any $n\ge N_0$ we have 
$\| \rho^{n,*}(T_1) - \bar\rho\|_\infty < \varepsilon \le \delta_0/(16)$.
We can thus apply Lemma \ref{inter} and get, for $n\ge N_0$
\begin{equation}\label{6}
I_1 \big(\hat{\pi}^{n,*} \big| \bar \rho \big) \le C
\Vert \rho^{n,*}(T_1) - \bar\rho \Vert_2^2 
\end{equation}
for some constant $C=C(\delta_0)$.

By Lemma \ref{st1}, $\rho^{n,*} (T_1-t)$, $t\in [0, T_1]$ 
is smooth and bounded away from $0$ and $1$, namely it belongs to
$D_{T_1,\delta_n}$ for some $\delta_n>0$. We can thus apply Lemma
\ref{I=I*+S} and conclude, as $\rho^{n,*}(T_1-t)$ solves 
\reff{f07bis} with $K=0$, 
\begin{equation}\label{7}
I_{T_1} \big( \rho^{n,*}(T_1-\cdot) \big| \rho^{n,*}(T_1)\big)
= S(\rho_n) - S(\rho^{n,*}(T_1)) \le S(\rho_n)
\end{equation}
From equations \reff{4}--\reff{7} we now get
$$
\begin{array}{rcl}
{\displaystyle 
I_T (\pi^* | \bar \rho) 
}
&\le & 
{\displaystyle 
\liminf_{n} \left[  S(\rho_n) + C \Vert \rho^{n,*}(T_1) - \bar\rho \Vert_2^2 
\right]
}
\\
&= & 
{\displaystyle \vphantom{\bigg(}
S(\rho) + C \Vert \rho^{*}(T_1) - \bar\rho \Vert_2^2 
\le S(\rho) + 2 C \e^2 
}
\end{array}
$$
and we are done by the arbitrariness of $\e$.

We are left to prove that $\pi^{n,*}\to \pi^{*}$ in $D\big([0,T], \mc
M\big)$. 
We show that $\pi^{n,*}$ converges to $\pi^*$ in $C\big([0,T];
\mc{M}\big)$.  Pick $\varepsilon_1 \in (0,T_1]$; since $\rho^{n,*}(t)$
converges to $\rho^{*}(t)$ in $C\big([-1,1]\big)$ uniformly for $t\in
[\varepsilon_1,T_1]$ we conclude easily that $\pi^{n,*}$ converges to
$\pi^*$ in $C\big([1,T-\varepsilon_1]\times [-1,1]\big)$.  We recall
that, by Lemma \ref{t:sch}, $\nabla F_n(t)$ and $\nabla F(t)$ are
uniformly bounded. Moreover, $\pi^{n,*}(T-t)$ and $\pi^{*}(T-t)$,
$t\in [T-T_1,T]$ are weak solutions of \reff{adjhyd}; for each $G\in
C\big([-1,1]\big)$ we thus get
$$
\lim_{\varepsilon_1\downarrow 0} \; \limsup_n \;
\sup_{t\in[T- \varepsilon_1,T]} 
\left| \langle\pi^{n,*}(t),G\rangle -\langle\pi^{*}(t),G\rangle \right|
=0
$$
this concludes the proof that $\rho^{n,*}$ converges to  $\rho^{*}$ in 
$C\big( [1,T];\mc{M} \big)$. Since $\rho^{*,n}(T_1)$ converges to 
$\rho^{*,n}(T_1)$ in $C^2\big( [-1,1]\big)$ it is easy to show that 
$\hat{\pi}^{n,*}$ converges to  $\hat{\pi}^{*}$ 
in $C\big( [0,1]\times[-1,1]\big)$. 
Hence $\pi^{n,*}$ converges to  $\pi^{*}$ in $C\big( [0,T];\mc{M} \big)$
\end{demo}

\appendix
\section{A lower bound on the quasi potential ($d\ge 1$)}
\label{s:app}

In this Appendix we prove a lower bound for the quasi potential in
the $d$ dimensional boundary driven simple exclusion process. 
For $d=1$ this bound has been derived from \reff{SsS} in 
\cite{DLSlet,DLS}. 

Let $\Lambda\subset \bb{R}^d$ be a smooth bounded 
open set and define $\Lambda_N := \bb{Z}^d \cap N\Lambda$.
Let also $\gamma(u)$ be a smooth function defined in a neighborhood of
$\partial\Lambda$.
The $d$--dimensional boundary driven symmetric exclusion process is
then the process on the state space $\Sigma_N:=\{0,1\}^{\Lambda_N}$ with
generator 
\begin{eqnarray*}
L_{N} f(\eta) &=& {\frac {N^2}{2}} 
\sum_{ 
\genfrac{}{}{0pt}{1}{\{x,y\}\subset\Lambda_N}{|x-y|=1} 
} 
\left[ f(\sigma^{x,y}\eta) -f (\eta)\right]
\\
&& \vphantom{\Bigg\{}+\;{\frac {N^2}{2}}
\sum_{
\genfrac{}{}{0pt}{1}{
x\in\Lambda_N, y \not\in\Lambda_N 
}
{|x-y|=1} 
}
\Big(
 \eta(x) + [1-\eta(x)] \gamma \big(\frac {y}{N} \big)
\Big) \left[ f(\sigma^x\eta) - f (\eta)\right]
\end{eqnarray*}
where $\sigma^{x,y}$ and $\sigma^x$ have been defined in Section
\ref{s:nr}.

The hydrodynamic equation is given by the heat equation in $\Lambda$,
namely 
$$
\left\{
\begin{array}{l}
\partial_t \rho = \frac 12 \Delta \rho \qquad u\in\Lambda \\
\rho (t,u) = \alpha (u) \qquad u\in\partial\Lambda \\
\rho(0,u) = \rho_0(u)
\end{array}
\right.
$$
where $\alpha(u) = \gamma(u)/[1+\gamma(u)]$. We shall denote by 
$\bar\rho=\bar\rho(u)$, $u\in\Lambda$ the unique stationary solution 
of the hydrodynamic equation.

By the same arguments as the ones given in Section \ref{s:3} it is
possible to prove the dynamical large deviation principle for the
empirical measure. The rate function is still given by the variational
formula \reff{eq:3}, but we now have
\begin{eqnarray*}
J_{T,H,\rho} (\pi) &:=& \big\langle \pi(T), H(T) \big\rangle 
- \langle \rho, H(0)\rangle
- \int_0^{T} \!dt\, \Big\langle \pi(t), \partial_t H(t) 
+ \frac 12 \Delta H(t) \Big\rangle
\\
&& -\frac{1}{2} \int_0^{T} \!dt\, 
\big\langle \chi( \pi(t) ), \big( \nabla H(t) \big)^2 \big\rangle \; 
+ \frac 12 \int_0^{T} \!dt \: \int_{\partial\Lambda} \!d\sigma(u) \: 
\alpha(u) \partial_{\hat n} H(t,u)
\end{eqnarray*}
where $\partial_{\hat n} H(t,u)$ is the normal derivative of $H(t,u)$
($\hat n$ being the outward normal to $\Lambda$) and $\sigma(u)$ is
the surface measure on $\partial\Lambda$.

Let us define the quasi potential $V(\rho)$ as in \reff{qp} and set 
$$
S_0(\rho) := \int_\Lambda \!du \: 
\left[ \rho(u) \log \frac {\rho(u)}{\bar\rho(u)} 
+ [1-\rho(u)] \log \frac {1-\rho(u)}{1-\bar\rho(u)} \right]
$$

\begin{theorem}
\label{qp>s_0} 
For each $\rho\in \mc M$ we have $V(\rho)\ge S_0(\rho)$.
\end{theorem}

\begin{demo}
We shall prove that $I_T(\pi |\bar\rho) \ge S_0(\rho)$ for any
$\pi(\cdot)$ such that $\pi(0)=\bar\rho$ and $\pi(T)=\rho$. 
Let us assume first that 
$\pi\in C^{1,2} ([0,T] \times \Lambda)$, $\pi(t,u)=\alpha(u)$ for 
$(t,u)\in [0,T]\times \partial\Lambda$, and $\pi$ is bounded away
from 0 and 1. Given such $\pi$ we use the variational characterization
of $I_T$ and chose 
$$
H(t,u)= \log \frac{\pi(t,u)}{1-\pi(t,u)} -
\log \frac{\bar\rho(u)}{1-\bar\rho(u)}
$$
Note that $H(t,u)=0$ for  $(t,u)\in [0,T]\times \partial\Lambda$
since $\pi$ and $\bar\rho$ satisfy the same boundary condition. 
By dominated convergence and an explicit computation we get
$$
S_0(\pi(T)) - S_0(\pi(0)) = 
\int_0^T  \!dt\:  \frac {d}{dt} S_0(\pi(t))
=  \int_0^T  \!dt\:  \left\langle \partial_t \pi (t), H(t) \right\rangle
$$

Recalling that $J_{T,H,\bar{\rho}}(\pi)$ has been defined above, a simple 
computation shows 
\begin{eqnarray*}
J_{T,H,\bar\rho} (\pi) &= & S_0(\pi(T)) + \frac 12 
\int_0^T \!dt\: 
\left\langle  \nabla H(t),  \nabla \pi(t)  -  \pi(t)[1-\pi(t)]  \nabla
H(t)\right\rangle 
\\
&=& S_0(\pi(T)) + \frac 12 
\int_0^T \!dt\: 
\left\langle \frac{(\nabla_u \bar \rho)^2}{[\bar\rho (1-\bar\rho)]^2}
\left( \pi(t) - \bar\rho \right)^2\right\rangle 
\end{eqnarray*}
since the second term above is positive  we conclude the proof of the
lemma for smooth paths. 
To get the general result it is enough to repeat the
approximation used in Lemma \ref{qp>s}.
\end{demo}

\section*{Acknowledgements}
We are grateful to G. Dell'Antonio for a useful discussion on the
variational problem which led to Lemma \ref{t:ddeq}.
We also thank T. Bodineau and G. Giacomin for stimulating discussions.

\end{document}